\documentclass[12pt,preprint]{aastex}
\shorttitle{Strong Mg II Absorption}
\shortauthors{Prochter \& Prochaska}

\begin{document}

\newcommand{\lya}{Ly$\alpha$}
\newcommand{\kms}{km~s$^{-1}$ }
\newcommand{\cm}[1]{\, {\rm cm^{#1}}}
\newcommand{\mkms}{{\rm \; km\;s^{-1}}}
\newcommand{\delv}{\Delta v}
\newcommand{\lofx}{$\ell_{\rm Mg}(X)$}
\newcommand{\lofz}{$\ell_{\rm Mg}(z)$}
\newcommand{\mlofx}{\ell_{\rm Mg}(X)}
\newcommand{\mlofz}{\ell_{\rm Mg}(z)}
\newcommand{\msol}{M_\odot}

\title{On the Incidence and Kinematics of Strong Mg II Absorbers}

\author{Gabriel E. Prochter, Jason X. Prochaska}
\affil{Department of Astronomy and Astrophysics, University of 
       California, 1156 High Street, Santa Cruz, CA  95064}
\email{prochter@ucolick.org,xavier@ucolick.org}
\and
\author{Scott M. Burles}
\affil{Kavli Institute for Astrophysics and Space Research \& Department 
of Physics, Massachusetts Institute of Technology, Cambridge, MA 
02139-4307, U.S.A.}
\email{burles@mit.edu}

\begin{abstract}
We present the results of two complementary investigations into the nature 
of strong (rest equivalent width, $W_r > 1.0$~\AA) \ion{Mg}{2} 
absorption systems at high redshift.  The first 
line of questioning examines the complete
Sloan Digital Sky Survey Data Release 3 set of 
quasar spectra to 
determine the evolution of the incidence of strong \ion{Mg}{2} absorption.  
This search resulted in 7421 confirmed \ion{Mg}{2} systems of $W_r > 1.0$~\AA\
yielding a $>95\%$ complete statistical sample of 4835 absorbers (systems 
detected in $S/N > 7$ spectral regions) spanning a redshift range $0.35 < z < 2.3$.
The redshift evolution of the comoving line-of-sight number density, 
\lofx\ is characterized by a roughly constant value at $z>0.8$
indicating the product of the number density and gas cross-section of
halos hosting strong \ion{Mg}{2} is unevolving at these redshifts.
In contrast, one observes a decline in \lofx\ at $z<0.8$ which we
interpret as a decrease in the gas cross-section to strong \ion{Mg}{2}
absorption and therefore a decline in the physical processes
relevant to strong \ion{Mg}{2} absorption.  
Perhaps uncoincidentally, this evolution
roughly tracks the
global evolution of the star formation rate density. 
Dividing the systems in $W_r$ sub-samples, the \lofx\ curves
show similar shape with lower normalization at higher $W_r$ values
and a more pronounced decrease in \lofx\ at $z<0.8$ for larger 
$W_r$ systems.
We also present the results of a search for strong 
Mg\,II absorption in a set of 91 high 
resolution quasar spectra collected on the ESI and HIRES spectrographs.  
These data 
allow us to investigate the kinematics of such systems at $0.8 < z < 2.7$.  
In this search a total of $22$ systems of $W_r > 1.0$~\AA were discovered.
These systems are characterized by the presence of numerous components spread 
over an average velocity width of $\delv \approx 200$~\kms. Also,
absorption due to more highly ionized species (e.g., Al\,III, C\,IV, Si\,IV) tend to 
display kinematic profiles similar to the corresponding 
\ion{Mg}{2} and \ion{Fe}{2} absorption.
We consider all of these results in light of two competing theories previously 
introduced to explain strong Mg\,II absorption:  post-starburst, supernovae-driven 
galactic winds and accreting gas in the halos of massive galaxies.  
The latter model is especially disfavored by the absence of evolution in 
\lofx\ at $z>1$.
We argue that the strong \ion{Mg}{2} phenomenon primarily arises
from feedback processes in relatively low mass galactic halos
related to star formation.
\end{abstract}


\section{Introduction}

Starbursting (SB) regions are likely to host rapidly expanding, highly ionized 
gaseous bubbles ("superwinds").  These bubbles are metal-enriched and powered by 
the correlated supernovae explosions of the stars on the high end of the SB
region's mass-function.  
If the starburst is related to a merger event, shock-heating during
the interaction may also lead to a wind of ionized gas
\citep{cox04}.
When nested in low enough mass galaxies, these bubbles
have a good chance of overpowering their host's potential well and escaping to 
pollute the intergalactic medium (IGM) 
\citep{FL03,MFR},
though \cite{cox04} note that even at high mass merging galaxies may have 
sufficient energy to produce expanding, gaseous bubbles.
At high redshift, such bubbles 
are expected to travel at velocities of 100 to 1000 $\textrm{km} ~\textrm{s}^{-1}$ and eventually expand 
to radii of $\sim 100 ~\textrm{kpc}$.
Presumably the winds entrain the metals of prior and recent supernovae
activity; therefore, the bubbles may cause metal-line
absorption in background sources in the $0.5 < W_r < 5$~\AA\ range.
Indeed, at low redshift the dwarf starbursting galaxy NGC~1569 
is observed to possess these characteristics \citep{MKH},
as is NGC 1705, another dwarf starburster \citep{HSM01}.  This 
enrichment process is believed to have played a leading role in the production 
of the observed metallicity of the present-day IGM
\citep{MFR,SFM}, and therefore further 
observational understanding of this pollution mechanism is key to 
the refinement of enrichment scenarios.
Furthermore, an exploration of the starbursting regions in absorption 
will constrain dynamic models of starburst winds and 
serve as an observational bridge for 
understanding the starbursting process at high redshift. 

A link has recently been proposed between these metal-enriched expanding 
bubbles and the very strong Mg II absorption systems 
($W_r > 1.8$~\AA) observed in quasar spectra
(\cite{BCCV}, hereafter BCCV).  In this paper the cutoff $W_r$ has been 
relaxed to $> 1.0$~\AA, as there is no known specific cutoff to the absorption 
depth of these systems and, as demonstrated later, smaller systems display 
features consistent with this phenomena.  It is, however, demonstrated 
that stronger systems are more likely to be associated with this absorption 
host scenario.
Assuming a simple spherical shape for the expanding shell, the expected profile 
for such a system is two roughly symmetric absorption lobes, corresponding to 
the cooler, expanding regions of the bubble, one in front and one behind the 
host galaxy along the line of sight, centered around a kinematically 
small, hot, non-absorbing region (a spectral "inversion"), where the actual 
host is located.  As the expanding 
gas is subject to Raleigh-Taylor instability, it is expected to clump, causing 
distinct kinematic components to appear in the absorption spectra.  
Cloud-to-cloud metallicity variation may also be expected, a result of 
unique supernova history and/or incomplete mixing.  
The expected velocity range of such systems 
is $\sim 100 - 1000 ~\textrm{km} ~\textrm{s}^{-1}$.  
BCCV identified all of these features in the profiles of a small set
of strong Mg\,II absorption systems.
Similarly, \cite{steidel02} find that gas "infall" from such winds 
could help explain the kinematics they observed in the spectra of a strong Mg II
system.  These authors also emphasized that
the study of more highly ionized species (e.g.\ C\,IV)
may further reveal the origin of the Mg\,II gas and its relation
to SN superwinds \citep[see also][]{bccv2}.

Based on the assumption 
that strong Mg II absorbers trace superwinds, and therefore starbursting 
galaxies, BCCV make a crude prediction of the redshift path 
density ($dN/dz$) of Mg II absorption.  These authors found
it to be consistent with a value 
extrapolated from the Mg II survey presented in \cite{SS92}, hereafter SS92,
though the comparison suffers from small number statistics.
Several goals of the work presented here include an 
extension of the work of BCCV to higher redshift, to obtain a 
large statistical sample,
and to investigate evolution with redshift.

An interesting correlation could also be made between strong Mg\,II 
absorption systems
and Lyman Break Galaxies (LBGs), which are thought to frequently harbor 
starbursting regions.
Comparisons of the velocities of the interstellar absorption
features with nebular emission lines 
suggests wind speeds of $200-1100$ \kms
\citep[e.g.][]{pettini01}.
In the gas, 
one identifies metal-line transitions of both low and high-ion species
with significant column densities. 
Therefore, one may probe the physical characteristics of these
winds (e.g.\ speed, cross-section, ionization state, etc.) through
absorption-line studies.  In turn, these observations should
constrain the incidence and evolution of superwinds during the first
$\sim 5$~Gyr of the universe.

A competing scenario for the origin of strong
Mg\,II absorption comes from  \cite{MM96}, hereafter MM96.
These authors describe a model 
where the absorption is associated with gaseous 
galactic halos in a two-phase structure: a shock heated hot phase and a 
photoionized phase of clouds moving through the halo.  The authors associate this 
phenomenon with Lyman limit systems, though note that at high redshift 
($z > 2$) the model predicts too few Lyman limits.  
An important aspect of this scenario is that strong Mg\,II absorption
is primarily restricted to the halos of relatively massive galaxies.
This is because the shock heated gas can cool only where  the gas is dense 
enough to radiate efficiently, limiting the possible extent of the cold gas.
Presumably the authors were motivated by observations showing that
strong Mg\,II systems at $z \lesssim 1$ are 
frequently identified at impact parameters less than $30 ~h^{-1}$~kpc
from $L \approx L^*$ galaxies \citep{lzt93,steidel93}.
In any case, 
the MM96 scenario ties Mg\,II absorption 
to massive, passively evolving galaxies rather 
than merging or starburst events. 

This paper presents two lines of investigation into the
nature of strong (equivalent width\footnote{Unless otherwise stated, $W_r$
refers to the equivalent width of the Mg\,II $\lambda 2796$ transition.}
$W_r > 1$~\AA) Mg\,II systems:
(1) an automated search of the quasar spectra database of the 
Sloan Digital Sky Survey Data Release 3 (SDSS-DR3) for Mg\,II systems
with $z=0.35 - 2.3$; and
(2) the analysis of a sample of Mg\,II systems ($z=0.8-2.7$) identified
in  high resolution spectra of 92 quasars.
The paper is organized as follows.  $\S$ 2 explains the data used in this 
study and the analysis undertook on it.  $\S$ 3 presents our results, 
including a plot of $dN/dz$ versus redshift, a tabulation of each of the 
discovered systems, and a plot of interesting metal absorption lines.  This 
section also includes a discussion of the relevance of 
our results to models of metal pollution of the IGM.  
The paper concludes in $\S$~4 with a few remarks about future avenues
of research.

\section{Data and Analysis}

This section describes two surveys for strong Mg\,II lines.
In the first, the moderate resolution spectra of Data Release~3
from the Sloan Digital Sky Survey are searched for absorption.
This survey provides the largest sample of Mg\,II systems compiled to date.
The second data set is a much smaller sample of high resolution spectroscopy
originally obtained to study high $z$ damped \lya\ systems.
This data extends the search to higher redshift and, more importantly,
allows a quantitative assessment of the kinematics of Mg\,II absorbers.

\subsection{SDSS-DR3}

The first set of data comes from Data Release~3 of the 
Sloan Digital Sky Survey (SDSS-DR3).   
The sample has been restricted
to SDSS-DR3 quasars with redshift $z_{qso} > 0.35$, for 
a total of 45023 quasars.  SDSS spectroscopy has a resolution 
$R \approx 2000$ and covers $3800-9200$~\AA.  This 
coverage implies a redshift search window of $0.35 < z < 2.3$ for the
Mg\,II~$\lambda\lambda 2796,2803$ doublet.

A search and analysis of strong Mg\,II 
absorption was carried out in several steps.  
First each spectrum was 
continuum fit red-ward of \lya\ emission
using a Principal Component Analysis to fit the quasar
emission lines and a b-spline algorithm to fit the underlying spectrum, which results
roughly in a power-law with broad emission features.
Second, all $3.5 \sigma$ features were identified using a Gaussian filter method
matched to the spectral resolution of the SDSS data.
This method works by convolving a Gaussian with $\sigma=1$\,pixels to both the 
flux and error arrays.  The wavelengths where the convolved flux exceeds the 
convolved error array by 3.5 are considered significant absorption features and a 
line list is constructed for each quasar.
Each resulting line list was searched for pairs of 
absorption features with separation matching the Mg\,II doublet.
Finally, the equivalent width of the Mg\,II~2796  
line was measured by directly summing the normalized flux. 
The equivalent width was measured by summing the difference between 
the continuum-fitted quasar flux and unity over a redshift centered box 
of width $3.589$~\AA\ (half the spectral distance between the rest 
Mg\,II(2796) and Mg\,II(2803) lines).  
This final step is dominated by systematic uncertainty (in particular 
continuum fitting and contamination due to absorption by both systems at 
other redshifts and the sky), with an expected typical error of 
$0.2 - 0.3$~\AA.
To remove false-positives, a visual confirmation was then made for each system with 
measured $W_r > 0.8$~\AA.  

\begin{figure}
\plotone{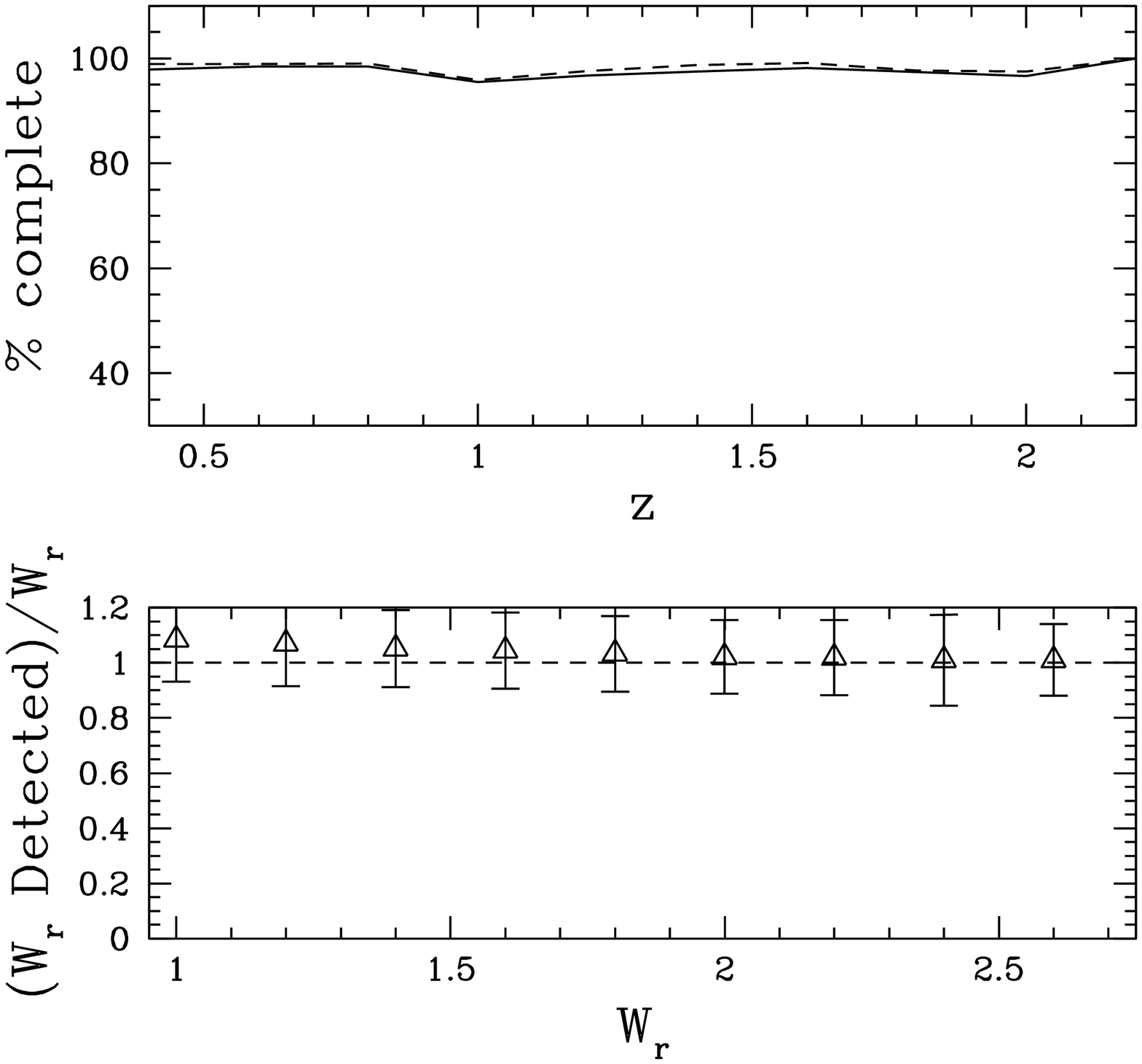}
\caption{Completeness measurements for the automated SDSS-DR3 absorption 
search, based on Monte-Carlo simulations ($S/N > 7$).  The top panel indicates 
percent completeness versus absorption redshift, with the solid line 
indicating systems of $W_r > 1.0$~\AA\ and the dashed line for systems of 
$W_r > 1.4$~\AA.  The bottom panel depicts the $W_r$ measured by the 
automated search algorithm versus the actual $W_r$ of the system added, with the 
vertical axis indicating the average ratio plus RMS error of systems for each 
0.1~\AA\ $W_r$ bin (the horizontal axis).
  \label{fig:completeness}}
\end{figure}

To check the completeness of the survey a Monte-Carlo test was run on 
a  representative sample of quasars.  Artificial Mg\,II systems were added to quasar 
spectra, which were then searched for absorption as described above.  The artificial 
systems were varied in strength and redshift and the quasars were selected at
random from the complete SDSS-DR3 
quasar sample.  The results of this simulation are displayed in 
Figure~\ref{fig:completeness}.  The solid line in the top panel reflect results 
for $W_r > 1.0$~\AA\ systems versus redshift.  The $W_r > 1.4$~\AA\ results are 
displayed with a dashed line.  The lower panel displays the $W_r$ of the systems 
as measured by the search algorithm versus the actual $W_r$ of the added system.
The data have been binned and the values displayed are the average detected $W_r$ 
divided by the actual value, with the errorbars indicating the RMS of the 
systems in each bin.  
This figure indicates that this technique of absorption detection 
is $> 95\%$ complete for $W_r > 1.0$~\AA\ systems, and provides 
fairly accurate measurements of $W_r$, with typical errors of $0.1-0.2$~\AA.  A very small number of 
outliers in measured $W_r$ were detected in this analysis, primarily a result of 
sky-line blending due to small errors in the detected system redshift.  This 
effect was observed in very few systems, and is ignored in the following analysis.
There is a slight systematic elevation of measured $W_r$ versus the added $W_r$, which is 
an effect of contamination by real absorption systems in a given spectra and is ignored 
in the following analysis.
To further assess the completeness and accuracy of the search algorithm, fifty 
random quasars were examined by eye.
In this sample the automated search missed 2 out of 15 systems,
each with $W_r < 0.7$~\AA;
the $W_r$ measurements of the automated search 
agreed well with this small sample of found systems.  
Because of Nyquist biasing, 
the number of discovered $W_r > 1.0$~\AA\ systems is expected to be inflated by $W_r 
< 1.0$~\AA\ systems included due to error in the measurement of $W_r$.  As the 
phenomena considered in this work have no specific cut-off in $W_r$, and as the 
effect is likely to be small, this error is neglected in the following analysis.

\begin{figure}
\plotone{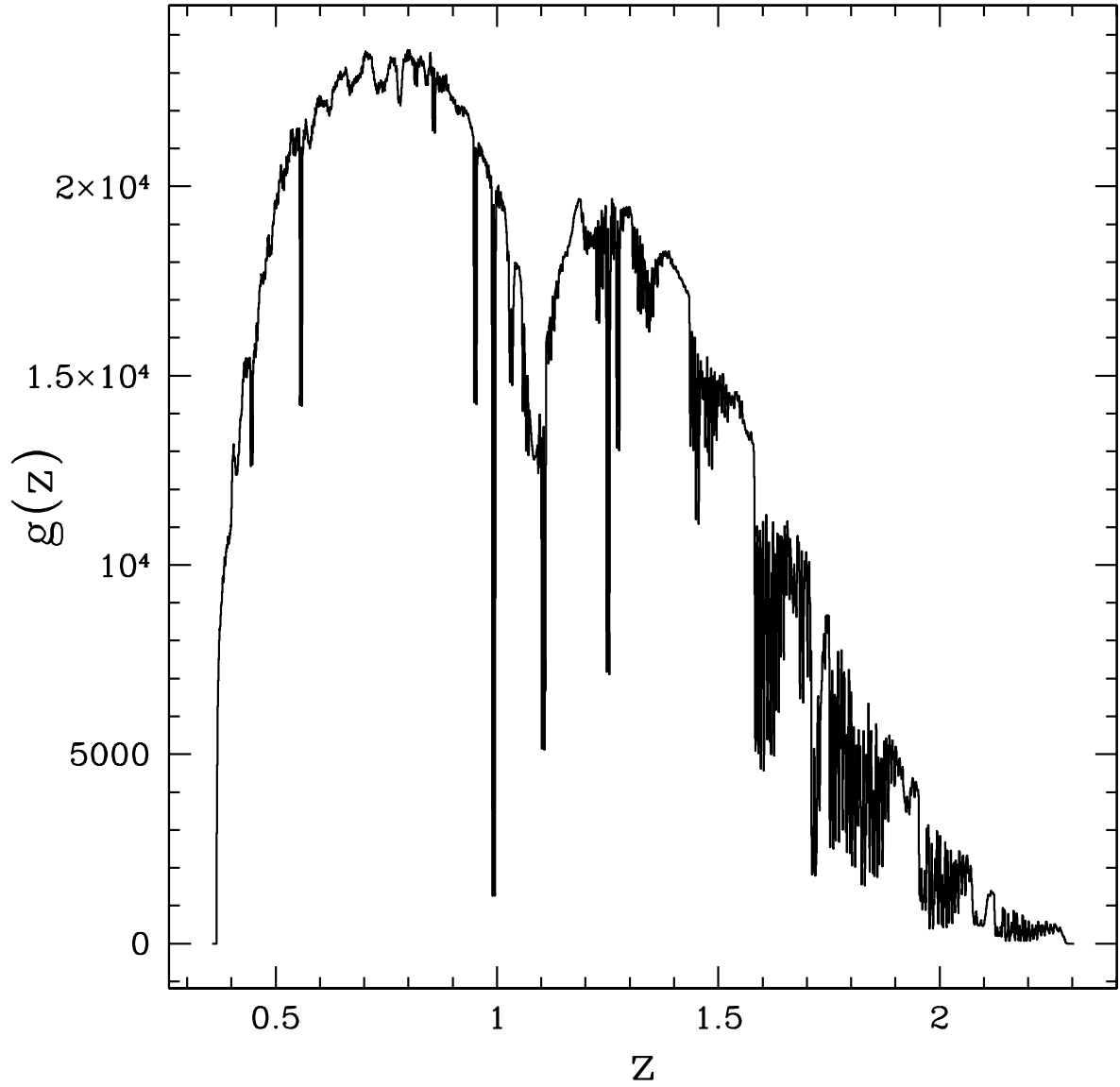}
\caption{The redshift path density, $g(z)$,  probed by SDSS-DR3.  
This represents the 
total number of lines 
of sight at a given redshift in which it is possible to detect
a strong Mg\,II system in the SDSS-DR3 spectra.  Clear in this plot are sky-absorption
features at high redshift, as well as very strong lines at $z \sim 1$ at $z \sim 1.25$.
  \label{fig:gzSDSS}}
\end{figure}

In order to determine the number density of Mg\,II systems, one must calculate
the total redshift path density $g(z)$ of the survey.  
Large spectral regions, especially at higher redshifts, are highly contaminated 
by the sky.  Simply removing these regions from the analysis would greatly 
reduce the redshift pathlength available to this study, but it is important to 
properly account for those regions which are too unreliable to accurately 
discover Mg\,II absorption.
To account for this issue, the search was limited to regions where the Gaussian fitted 
($\sigma$ = 1~pix) signal-to-noise (of the continuum) $S/N > 7$ per pixel, often
resulting in non-contiguous search regions for a given quasar.
The analysis of SS92 has been followed with a slight modification.
For this work, instead of adding unity to a $g(z)$ redshift 
bin for each spectrum which contains data for that bin, 
the decimal percentage of the given bin's 
redshift range actually covered
in each spectra is added.  
Because strong Mg\,II systems dampen the noise of a 
spectra at their wavelength (i.e.\ Poisson noise 
is a significant source of error), 
and the $S/N$ is calculated by dividing the 
continuum flux by the noise, the $S/N$ at the specific wavelength of a given 
system will tend to increase over the local average.  To protect against 
over-counting Mg\,II systems versus $g(z)$, only systems discovered 
within regions in which the $S/N$ smoothed over a 50 pixel box was greater than 
7 are considered for the following analysis.
A plot of $g(z)$ for the SDSS search is presented in Figure~\ref{fig:gzSDSS}.
Limiting the search to smoothed spectral regions of $S/N > 7$ per pixel
and a minimum $W_r$ threshold of $0.8$~\AA, 
a total of 6511 Mg\,II systems were discovered, with 4835 of 
$W_r > 1.0$~\AA, 2594 of $W_r > 1.4$~\AA\ and 
1329 of $W_r > 1.8$~\AA.  Another 3031 systems were discovered outside of 
$S/N > 7$ regions, 2586 of which have $W_r > 1.0$~\AA.  

\begin{figure}
\plotone{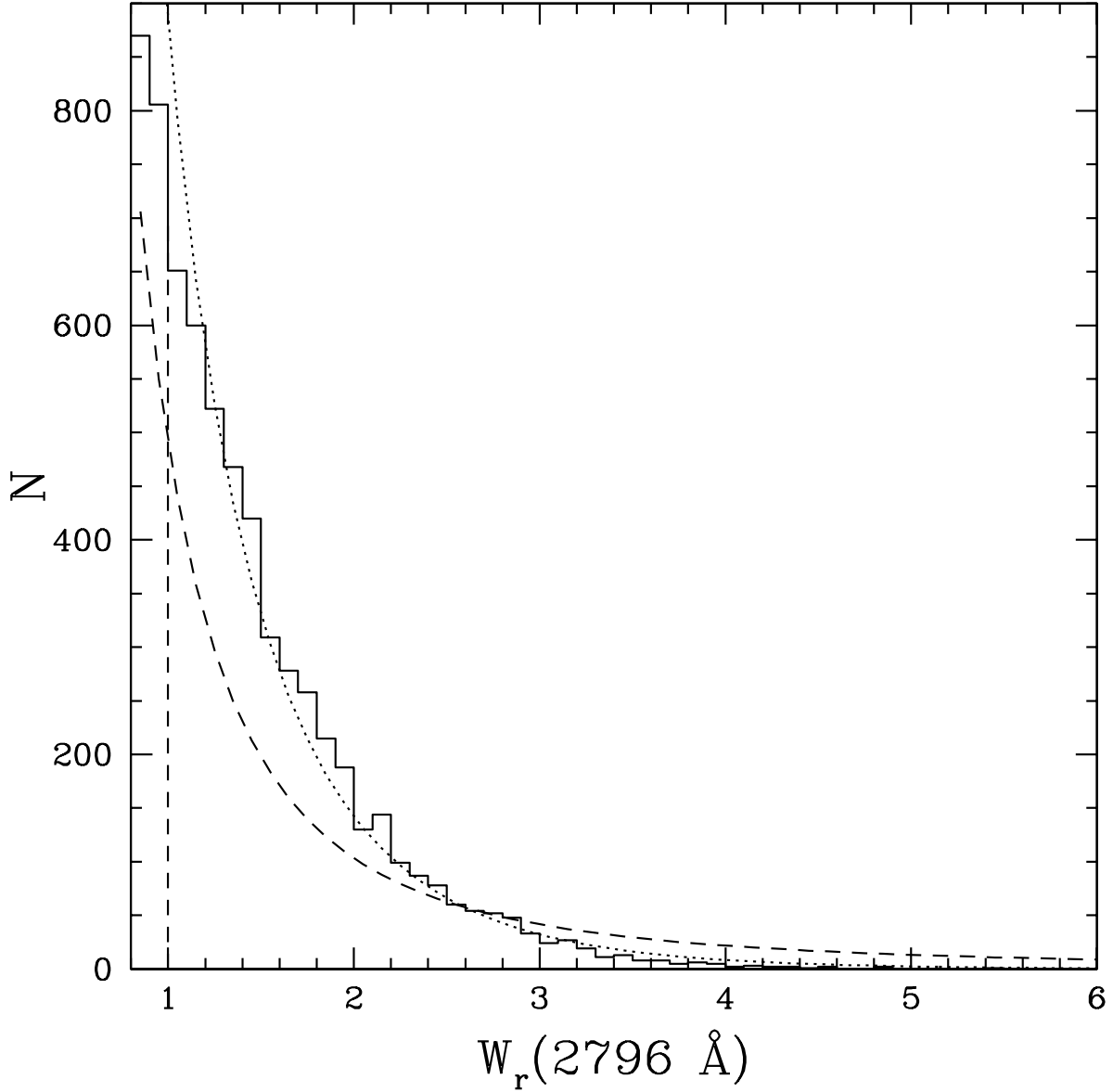}
\caption{A histogram of the $W_r$ values 
for the Mg\,II systems discovered in SDSS-DR3
with our automated search algorithm.  The vertical
  dashed line is located at $W_r = 1.0$ \AA. 
Given that the incidence of $W_r$ continues to rise well below 1\AA,
the survey is expected to be $>95\%$ complete.
A power law fit for systems of $W_r > 1.0$\AA\ is plotted as a dashed line;
$f(W_r) = 490.37 W_r^{-2.245}$.  The dotted line represents a modified 
Schecter functional fit the the $W_r > 1.0$\AA\ data, $f(W_r) = 2398.4 
W_r^{-1.19} e^{-W_r}$.
  \label{fig:ewSDSS}}
\end{figure}

A histogram of $W_r$ values for the SDSS-DR3 sample is plotted in 
Figure~\ref{fig:ewSDSS}.
The distribution is steeply rising at $W_r = 1$~\AA, inspiring
confidence that the analysis is reasonably complete above this limit.
A power-law fit to the $W_r > 1.0$\,\AA\ distribution, 
$f(W_r) = CW^{-\delta}$, results in the 
parameters $C = 490.4(^{+8.3}_{-9.4})$ and 
$\delta = 2.24(^{+0.04}_{-0.04})$. 
This fit is plotted 
as the dashed line in Figure~\ref{fig:ewSDSS}. 
The slope of this function 
is steeper than that measured 
in SS92 ($\delta = 1.65\pm0.03$), but is more consistent
with the measurements of others \cite[][$\delta = 2.2\pm0.3$]{tytler87}.
It is clear from the figure, however, that
a power-law is not a good description of the observed distribution, 
under-predicting the number of systems at low $W_r$ and over-predicting 
the number of larger systems.
The dotted line plotted in the figure represents a fit of a modified 
Schecter function to the $W_r$ distribution, of the form
$f(W_r) = BW_r^{-\phi}e^{-W_r}$, which is a much better fit to the data, 
particularly at high $W_r$.  The best fit values for this functional form 
are $B = 2398(^{+72}_{-71})$ and $\phi = 1.19 \pm 0.07$.  It is likely that 
allowing the characteristic $W_r$, $W_r^*$, to be a fitted variable rather 
setting it to unity would improve the fit even further.  Because this value 
has no bearing on the following discussion, this fitting and its interpretation 
are left to future work.  All the fits were preformed using unbinned, Maximum
Likelihood analysis.  The errors are $95$\% confidence limits. 

Table~\ref{tab:sdss} presents the SDSS quasar name, RA and DEC, emission
redshift, the Mg\,II redshift, 
and $W_r$ of each $W_r > 1.0$~\AA\ 
Mg\,II system discovered in the automated SDSS search, 
including those which passed 
the 'by-eye' confirmation but were eliminated from further analysis by the $S/N > 7$ per pixel criterion.  

\subsection{High Resolution Observations}

The second set of quasar spectra used in this study includes 51
high $S/N$ spectra ($FWHM \sim 6.3 - 8.4 ~\textrm{km} ~\textrm{s}^{-1}$) 
taken with the High 
Resolution Echelle Spectrometer \citep[HIRES;][]{vogt} on the Keck I 10m 
telescope.  These data were recorded and reduced as part of the UCSD/Keck I Damped
Ly$\alpha$ Abundance Database and a full discussion of 
the reduction process can be 
found in Prochaska et al. (2001).  The high resolution sample also includes
41 $R \sim 8000$ spectra taken on the ESI spectrograph on Keck II.  
These data were collected as a part of the ESI/KECK II Damped \lya\ Abundance 
Database, details of which can be found in \cite{pro03}.
Table~\ref{tab:high} summarizes the spectra searched in this study, 
presenting the name of each quasar, $z_{em}$, data collection instrument
(with 1 and 2 referring to HIRES and ESI, respectively), the lower and upper 
limits of redshift available for Mg II detection, and information on discovered
absorption systems. 
It should be noted that every spectrum in this data set contains a 
damped \lya\ system;  these were ignored throughout the Mg\,II analysis.

For each spectrum a 'by-eye' search was made for Mg II absorption systems.  
Given the strength of this doublet, these systems are 
conspicuous.  
After identifying a Mg\,II system,
its redshift, equivalent width (Mg\,II 2796) 
and velocity width were measured, and the spectrum was 
searched for absorption due to other transitions 
associated with the system.  The 
velocity width ($\delv$) 
calculation followed the analysis of \cite{PP97}, 
i.e., the region bounding 90$\%$ of the total optical depth in
an unsaturated transition (typically Fe II 2600).
The kinematic complexity of each system was also assessed, generally using 
complementary, non-saturated metal absorption lines (e.g. Fe\,II 2600).  This 
measure is a simple summation of the distinct velocity components visible 
and was made due to the importance of absorption complexity to the arguments
of BCCV.  Because the ESI spectra are of lower resolution than 
those taken with HIRES, the observed complexity is expected to be reduced.  
Velocity structure will be lost for components separated by less than
45~\kms, especially in saturated absorption features.
On the other hand, the spectra are sensitive to
structure in the velocity profiles for components separated by greater
than $\approx 50 \mkms$.

\begin{figure}
\plotone{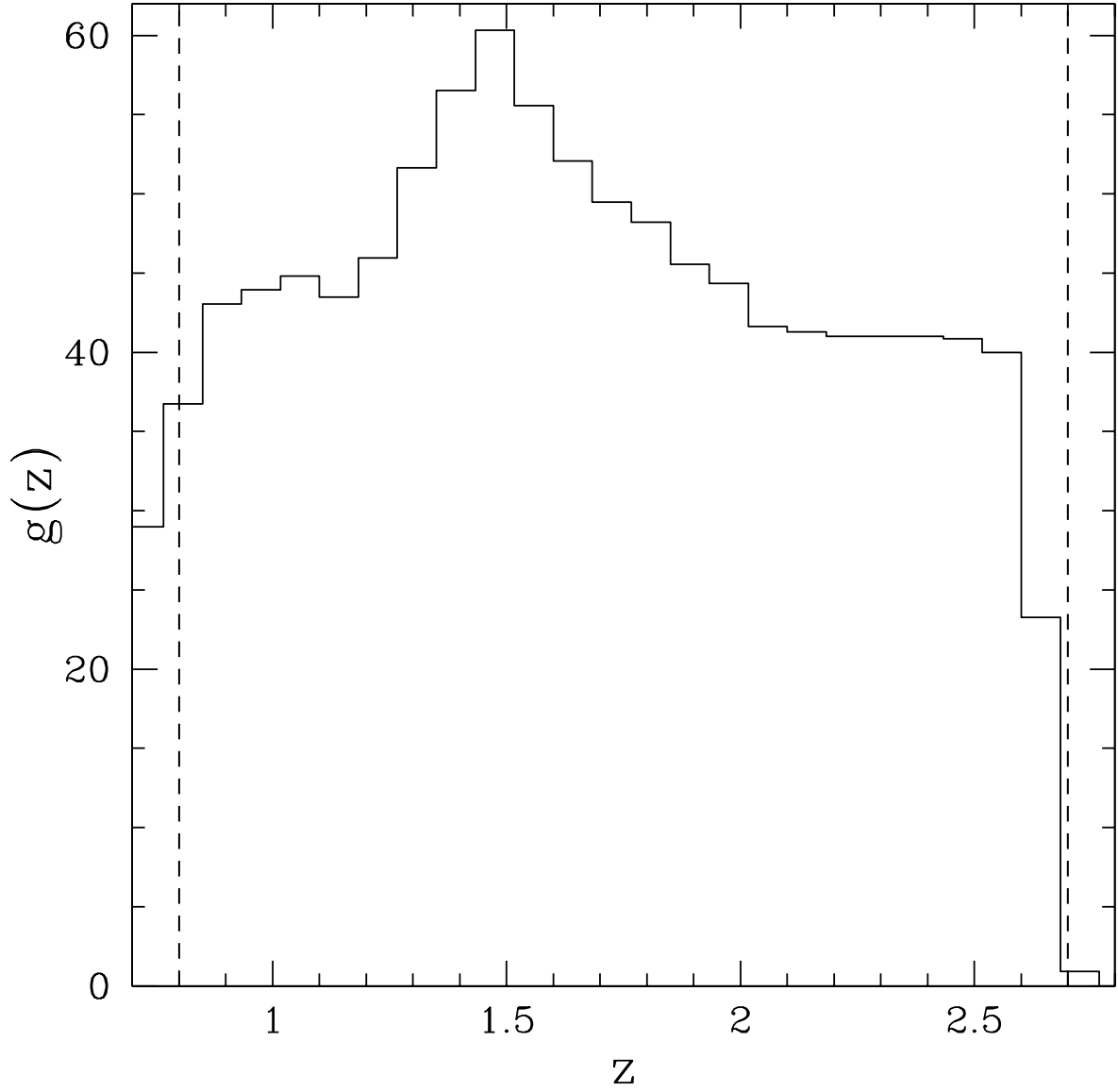}
\caption{The redshift path density, $g(z)$,  probed in the high resolution survey.
  The dashed vertical lines indicate the redshift region considered in the work
  presented in this paper.  \label{fig:gzhigh}}
\end{figure}

The redshift path density $g(z)$ has been calculated in a similar manner as
the SDSS sightlines.  The only significant difference is that
spectral gaps inherent to the HIRES instrument add some uncertainty
to this calculation.  A possible prescription for 
accounting for the HIRES spectral gaps involves disqualifying those spectral 
areas immediately preceding and following the gaps, as only one half of the 
doublet is not enough to determine its nature.  Because the systems of 
interest in this work are quite strong and because the wavelength coverage allows for 
investigation and confirmation of systems based on other metal absorption lines 
(e.g. Fe II),
this prescription was not followed and only the HIRES spectral gaps large 
enough to hide both components of a Mg II absorption doublet 
were disqualified in the calculation of $g(z)$.  
Figure~\ref{fig:gzhigh} presents $g(z)$ for the high resolution sample.  Clearly 
this high-resolution data extends to a higher redshift than available in the 
SDSS sample, though the small number of sightlines limits the significance of 
statistical results gleaned from this data.

\begin{figure}
\plotone{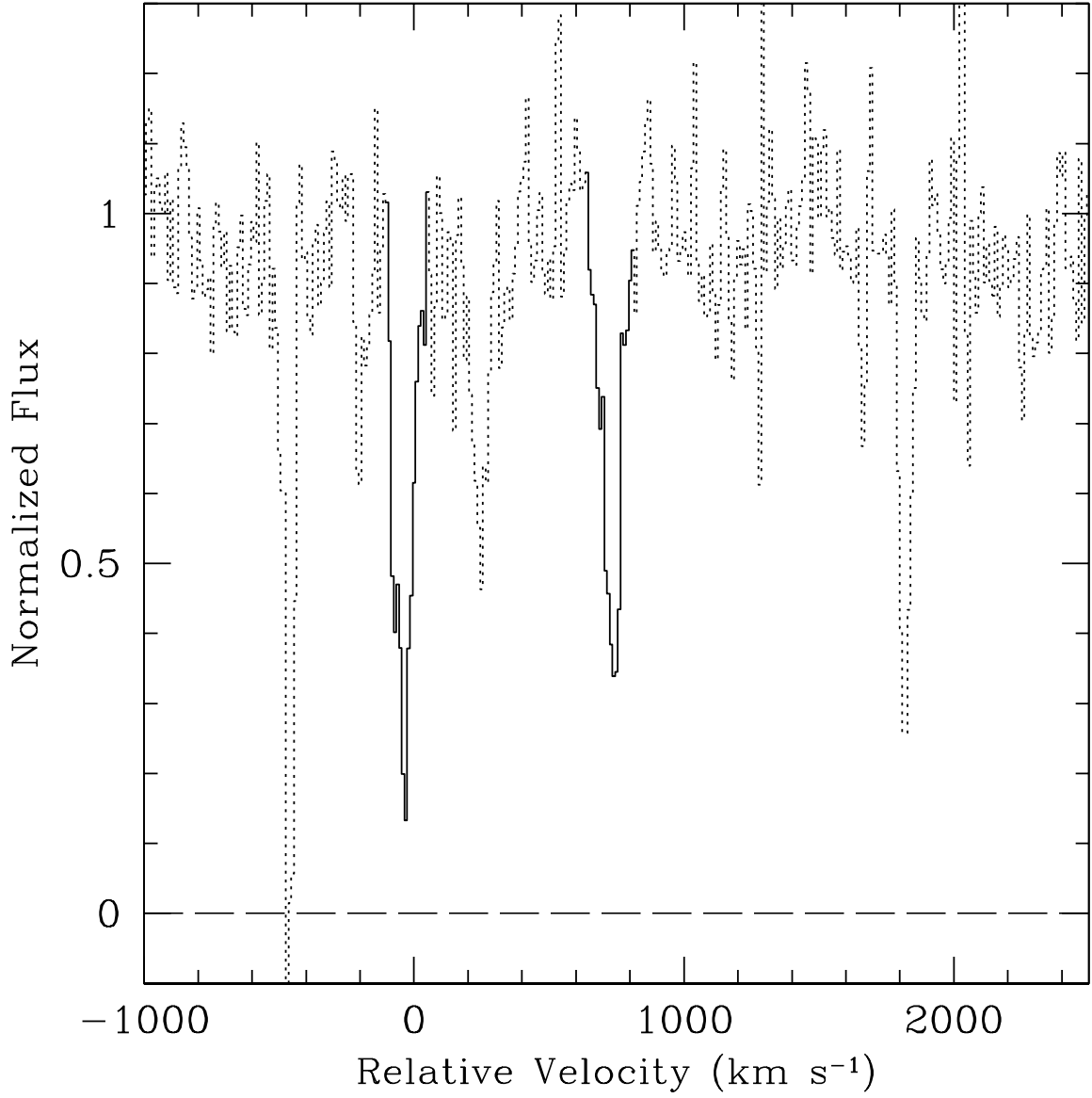}
\caption{An example Mg\,II system discovered 
at $z = 2.51$ in the spectrum of Q1337+11.
This system is located at $\sim 9800$\AA, where the 
sky begins to crowd the spectrum yet is easily 
identified at this spectral resolution and signal-to-noise.
  \label{fig:exsys}}
\end{figure}

The high resolution survey is 
complete to below the $W_r = 1.0$~\AA\ level.  
Both the ESI and HIRES 
datasets provide more than enough resolution and $S/N$ for confidence 
in this statement.  More than 50
Mg\,II systems with $W_r < 1$~\AA\ were discovered in the search,
though disregarded for the work presented here.  
At very long wavelengths ($\sim 10000$~\AA\ in the ESI spectra), 
sky lines can confuse the search for Mg\,II systems.
In a few particularly bad cases data are disqualified for this analysis, 
though the majority of spectra remained clear enough to easily identify 
absorption systems of the strength considered in this work.  An example 
is presented in Figure~\ref{fig:exsys}: a plot of the Mg\,II system 
with $W_r = 0.56 ~\textrm{\AA}$ located at $z = 2.51$ 
($\sim 9800 ~\textrm{\AA}$)
in the spectrum of Q1337+11.  Even crowded by the sky, this system is easily 
identified though it is only half as strong as the 
systems of interest to this study.

\begin{figure}
\plotone{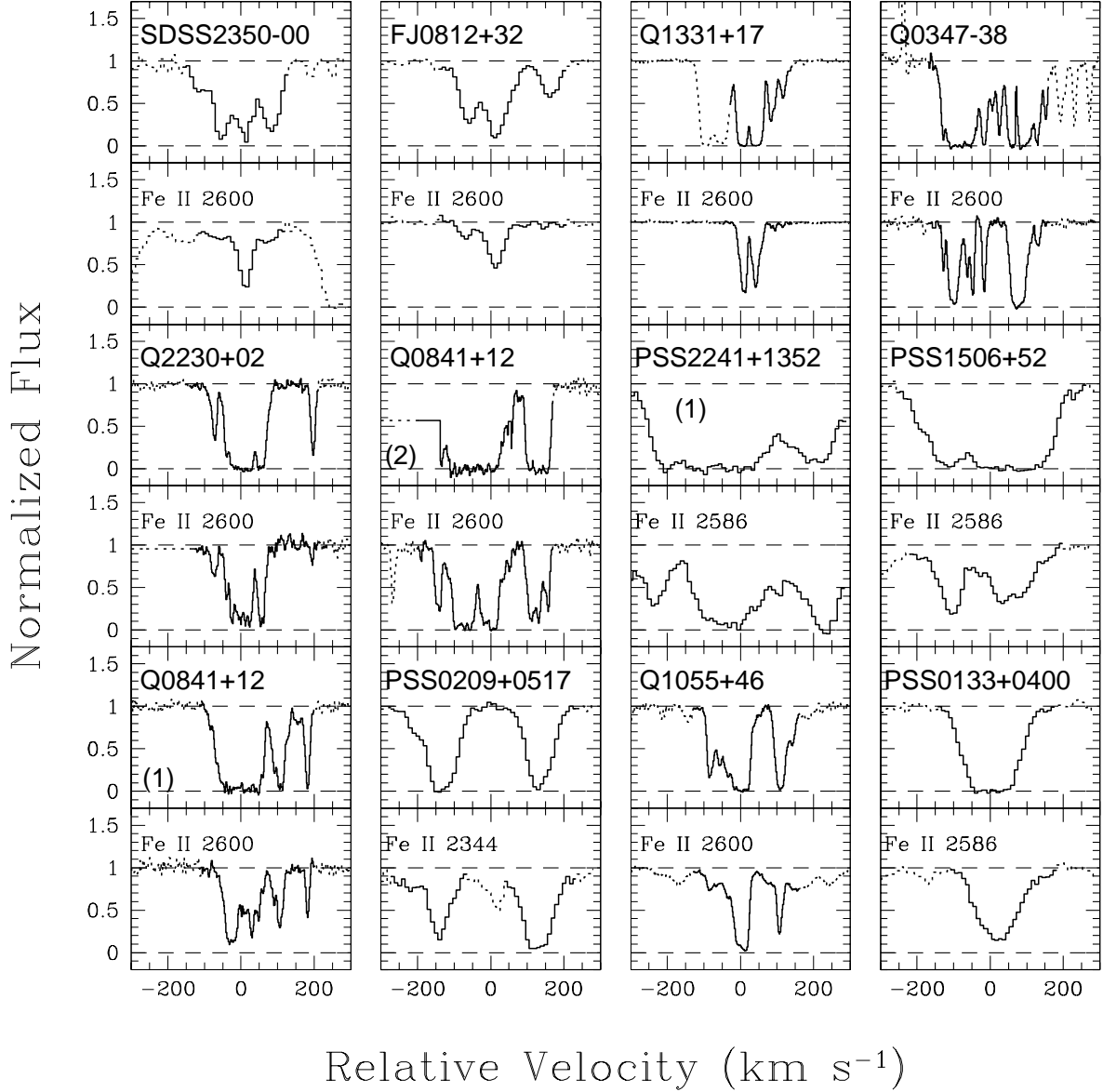}
\caption{The 22 detected $W_r > 1.0 ~\textrm{\AA}$, Mg\,II absorption systems
  detected in the high resolution survey.  For each system the Mg\,II 2796 
absorption line is displayed in relative velocity space 
with 0 $\textrm{km} ~\textrm{s}^{-1}$ centered at the 
  redshift reported in Table~\ref{tab:high}, with (1) referring to the low-z and (2) to 
  the high-z system in cases of multiple detections per quasar.  
  For each system, an associated, less saturated metal transition is displayed 
  below the Mg\,II 2796 feature.  The systems are ordered by increasing 
  redshift from top to bottom then left to right.
   \label{fig:data}}
\end{figure}

The 22 $W_r > 1.0 ~\textrm{\AA}$ Mg\,II absorption systems 
discovered in this survey are 
depicted in Figure~\ref{fig:data}.  For each system the 
Mg\,II 2796 absorption line is 
displayed in relative velocity ($\textrm{km} ~\textrm{s}^{-1}$) space with 0 
$\textrm{km} ~\textrm{s}^{-1}$ centered on the 
redshift reported in Table~\ref{tab:high}.  When more than one system was discovered
in a given quasar spectrum, the lower and higher redshift systems are labeled with 
a (1) and (2), respectively (no more than two systems were detected in any one 
spectrum).  An unsaturated or mildly saturated
transition (usually an Fe II line) is also displayed for 
each system. 

Six ($\sim$ 50\%) of the systems for which spectra at the 
relevant wavelength were available
(both observed in the quasar spectrum and not lost to the Ly$\alpha$ forest) 
contained 
evidence of absorption due to Si IV, C IV, and Al III.  
These lines are displayed in Figure~\ref{fig:ions},
where the systems are labeled as in Figure~\ref{fig:data}. 
These lines are displayed in relative velocity space with 0 
$\textrm{km} ~\textrm{s}^{-1}$ 
centered at the system redshift found in Table~\ref{tab:high}.

\section{The Incidence of Strong Mg\,II Absorption}

\subsection{Observations of and Fits to \lofx}

One means of investigating the physical nature of strong Mg\,II systems
is to examine their incidence as a function of cosmological time.
The most convenient and least ambiguous measure of this incidence is
\lofx, the line-density of absorbers per comoving Mpc.  
The incidence of a given class of absorbers at any redshift is proportional 
to their covering
fraction on the sky.  
If the \ion{Mg}{2} systems are a non-evolving population (i.e.\ constant
number per comoving Mpc$^{-3}$ $n_0$ and constant cross-section $\sigma_0$), 
then one would predict $\mlofx = n_0 \sigma_0$.
Should the number density or cross-section evolve with redshift, then 
$n_0$ and $\sigma_0$ are replaced with $n(z)$ and 
$\sigma(z)$.  If, for example, strong Mg\,II systems are related to merger 
events, then one would expect $n(z)$ to track the merger rate, predicted to be
proportional to $(1+z)^3$ \citep[e.g.,][]{merger}, which implies a steep
redshift dependence for \lofx.
Furthermore, because the number density evolution of halos in 
hierarchical cosmology is a sensitive function of mass, making 
comparisons with the evolution of \lofx\ may indicate the mass 
of the systems giving rise to \ion{Mg}{2} absorption. 

Historically, analysis has focused on the incidence of \ion{Mg}{2}
systems as a function of redshift, \lofz ~(frequently written as $dN/dz$).
Interpretation of this measure is somewhat ambiguous, 
because the evolution is influenced by the expansion of the universe
as well as any evolution of the absorbers in question

\begin{equation}
\mlofz = \mlofx * (c/H_0)(1+z)^2[\Omega_M(1+z)^3+\Omega_\Lambda]^{-1/2} \; .
\label{eqn:dndz}
\end{equation}

\noindent For this reason the majority of the 
analysis below is presented in terms of \lofx, which renders more obvious the 
trends discussed.  For comparison to previous work, statistics of \lofz\ are 
also presented.

\begin{figure}
\plotone{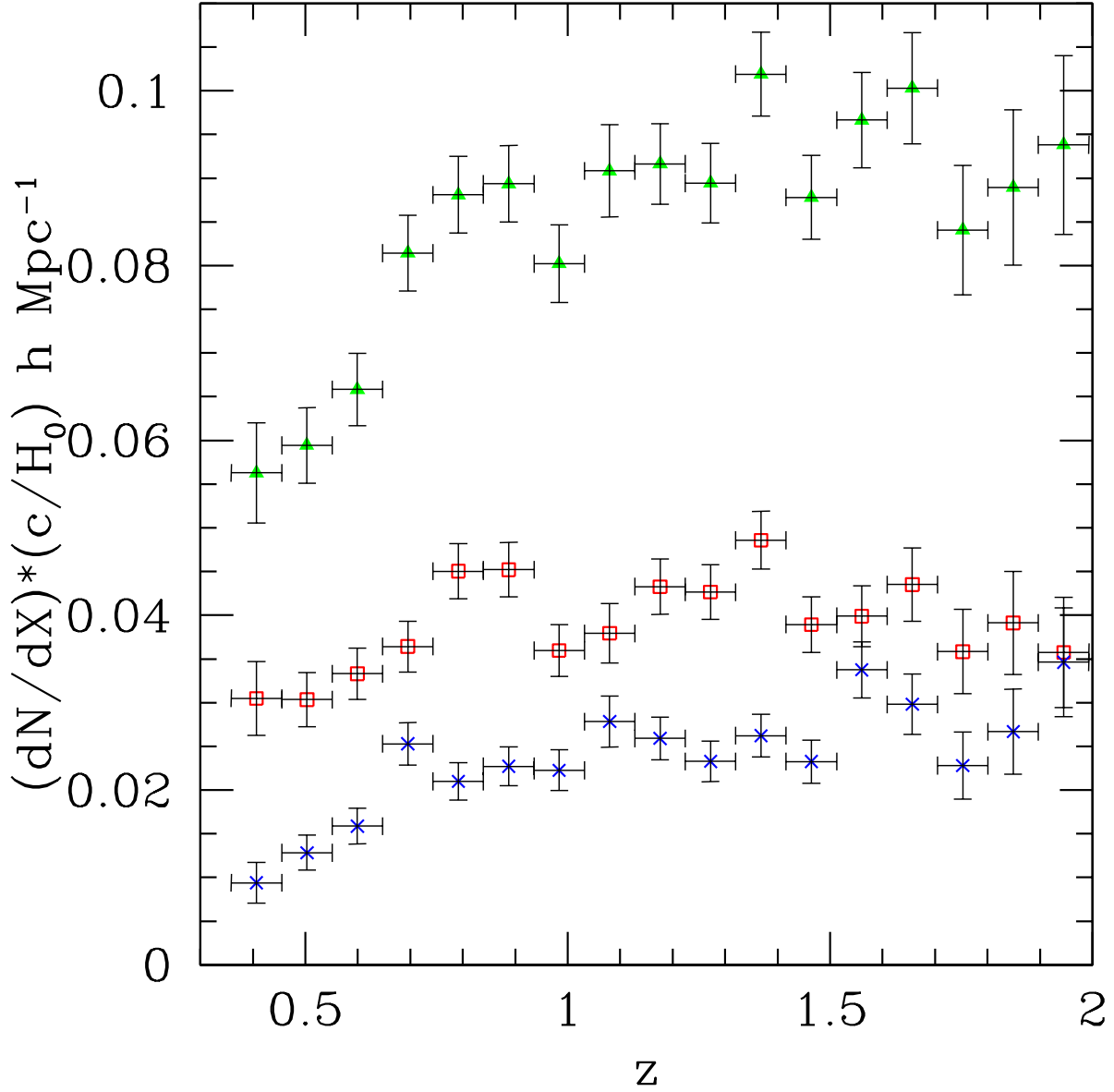}
\caption{The incidence of strong \ion{Mg}{2} absorbers, \lofx, 
from the SDSS-DR3 over the range $0.35 < z < 2$.
The triangles show the full sample $(W_r > 1$\AA),
the open squares are all systems of $1.0$~\AA\ $ < W_r < 1.4$~\AA\
and systems of $W_r > 1.8$~\AA\ are indicated with cross-hatches.  
The differences between the datasets
in this plot have important implications for the nature of Mg\,II absorption hosts.
   \label{fig:dndzSDSS}}
\end{figure}

Figure~\ref{fig:dndzSDSS} presents \lofx\ for
the Mg\,II systems identified in the SDSS-DR3 sample over the 
redshift range $z=0.35-2.3$.   The vertical errorbars 
assume Gaussian statistics while the
horizontal errorbars indicate the size of each redshift bin.  The SDSS results
are displayed in three sets:  
the closed triangles represent the entire $W_r > 1.0$~\AA\ dataset,
the open squares indicate systems of $1.0$~\AA\ $< W_r < 1.4$~\AA, 
and the cross-hatches represent $W_r > 1.8$~\AA. 
The data plotted in this figure are also listed in Table~\ref{tab:dndz},
as are the results for several other samples in $W_r$.
Qualitatively, the samples shown in Figure~\ref{fig:dndzSDSS}
exhibit similar \lofx\ evolution.
At redshifts $z>0.8$, \lofx\ is nearly constant and linear fits
yield slopes consistent with zero at the 95$\%$ c.l.
At $z< 0.8$, in contrast, each sample exhibits a decrease in \lofx.
For the $1.0$~\AA\ $< W_r < 1.4$~\AA\ sample the decline is 
marginally significant ($\sim 1\sigma$), while the effect is highly
significant for any sample with $W_r > 1.4$~\AA.

We have considered two functional forms to describe the
evolution of \lofx\ with redshift: (1) a linear fit $\mlofx = b + mz$
and (2) an exponential fit $\mlofx = N \exp(-z_0/z)$.
Fitting a line to the $1.0$\AA~$< W_r < 1.4$\AA\ sample yields
a slope consistent with zero: $m=0.002 \pm 0.003$.  In contrast,
the strongest \ion{Mg}{2} absorbers ($W_r > 1.8$\AA) 
have $m=0.011 \pm 0.002$.  Therefore, there is steeper increase
in \lofx\ for the larger $W_r$ absorbers, with the majority of
the difference occurring at $z<1$.
These functional forms have very large reduced $\chi^2$ values,
however, and are not good descriptions of the data.
The exponential fits, in contrast, are a reasonable model of the
data as these allow for
a sharp dropoff of incidence at low redshift.  Using standard
maximum likelihood methods, we find
$N = 0.044 \pm 0.003$ and $z_0 = 0.12 \pm 0.07$ for the 
$1.0$~\AA\ $< W_r < 1.4$~\AA\ systems, $N = 0.037 \pm 0.003$ and $z_0 = 0.47 
\pm 0.10$ for the $W_r > 1.8$~\AA\ systems, and $N = 0.11 \pm 0.006$ and 
$z_0 = 0.28 \pm 0.05$ for the full $W_r > 1.0$~\AA\ dataset.  
These results are summarized in Table~\ref{tab:errors}.

For comparison to previous work, power-law fits to the \lofz\
values of the form  $N_0 (1+z)^{\gamma_0}$ have also been performed. 
We stress, however, that such fits are no longer a good description
of the data.
In particular, this functional form cannot describe both the increase
in \lofx\ at low redshift and the roughly constant value at $z=1$ to 2.
Nevertheless, this analysis was done for several cuts of the data.  
The maximum likelihood values for $N_0$ and $\gamma_0$ are
$N_0 = 0.084, 
\gamma_0 = 1.40  $ for $W_r > 1.0$~\AA,
$N_0 = 0.051, 
\gamma_0 = 0.99$ for 1.0\AA~$W_r > 1.4$~\AA,
and 
$N_0 = 0.016, \gamma_0 = 1.92$ for $W_r > 1.8$~\AA.  
Table~\ref{tab:errors} reports central values for $N_0$ and $\gamma_0$, 
as well as 95\% confidence limits for each of the three fits.  
It should 
be noted that the values of $\gamma_0$ and $N_0$ are highly correlated.
Previous work on Mg\,II absorption has yielded similar, if somewhat steeper, 
results.  SS92 find that for $W_r > 1.0 $~\AA\ systems, 
$dN/dz \propto (1+z)^{2.24 \pm 0.76}$.
Also consistent with the results of SS92 is the increase in $\gamma_0$ 
with minimum equivalent width.  
We emphasize, however, that the differences in $\gamma_0$ are dominated
by the evolution in \lofx\ at low redshift because all of the
sub-samples have nearly constant \lofx\ at $z>1$.
Altogether, the $W_r > 1.0$~\AA\ sample has an average incidence
per unit redshift
$\mlofz = 0.174 \pm 0.002$ for the redshift region
$0.35 \leq z \leq 2.3$.  
This value is consistent with but $25\%$ lower than the value
reported by SS92 for the redshift range $0.2 \leq z \leq 2.2$ and
are well matched to those of \cite{sdssedr}, 
who report results from the SDSS Early Data Release.   

\subsection{Selection Bias}

It is important to 
understand and constrain possible biases in the SDSS-DR3 quasar sample.
The value of \lofx\ is sensitive to observational bias in several forms.
For example, if the Mg\,II systems are themselves massive enough to 
gravitationally lens, and thus brighten, their background quasars then 
the SDSS-DR3 would be statistically biased toward observing quasars 
which exhibit strong Mg\,II absorption \citep{menard05}.  This would lead to 
an underestimation of $g(z)$ and an overestimation of \lofx.  The opposite 
effect may result if quasar spectra are reddened by dust in the \ion{Mg}{2}
host systems.  Indeed, this bias is of particular concern, as many of the 
systems in this sample will be associated with damped \lya\
systems \citep{nestor03} and such systems may cause 
reddening in the spectra of background quasars \citep{pei91,murphy04}.

Following the analysis of \cite{murphy04} to address the matter of 
gravitational lensing, two averaged bootstrap 
samples of quasars free of strong Mg\,II absorption were created.  
The first of these samples was free of Mg\,II absorption in the 
range of $1.0$~\AA\ $< W_r < 1.4$~\AA, the second of $W_r > 1.8$~\AA\ 
systems.  These 
quasars where chosen to reproduce the redshift distribution of those 
sets of Mg\,II absorption-hosting quasars, but were otherwise chosen at random. 
Repeating this process many times uncovered the average non-absorption 
quasar magnitude distribution, which was then compared to the magnitude 
distribution of those quasars with Mg\,II absorption.  This analysis 
revealed no signal of gravitational lensing in either the $1.0$~\AA\
$< W_r < 1.4$~\AA\ or the $W_r > 1.8$~\AA\ datasets.  Indeed, two-sided 
Kolmogorov-Smirnov tests reveal no evidence of differing magnitude 
distributions between the absorption and non-absorption datasets in either 
$W_r$ bin, with KS probabilities ranging from $0.2-0.6$.

\begin{figure}
\plotone{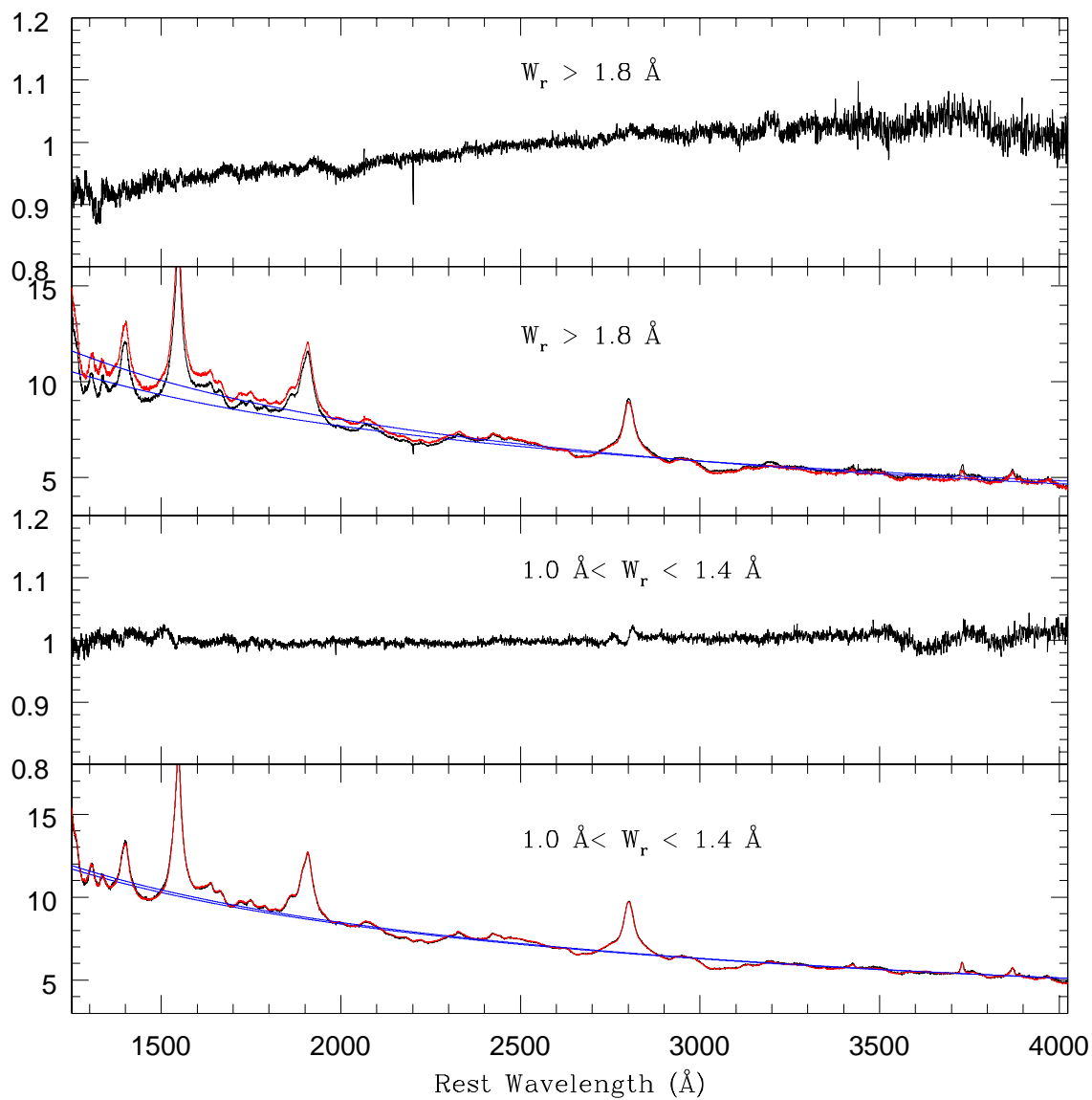}
\caption{Results from an analysis of the dust reddening of quasar spectra 
hosting strong Mg\,II absorption.  
The top two panels are results from the $W_r > 1.8$~\AA\ systems with the 
lower panel displaying composite spectra of quasars hosting strong Mg\,II 
absorption (black) and quasars without absorption (red).  The top panel 
is the absorption spectra divided by the non-absorption spectra.
The bottom two panels are the equivalent results for systems 
of $1.0$~\AA\ $< W_r < 1.4$~\AA. These results are indicative of the 
spectra of those quasars hosting the largest of Mg\,II systems having been 
reddened.
\label{fig:dust}}
\end{figure}

Dust however, does seem 
to affect the spectra on average of those quasars hosting the strongest 
of the Mg\,II absorption systems.  By building 
samples of quasars similar to the controls used in the lensing test (but 
now also constraining magnitude distributions of the 
control samples), composite `non-absorption' quasar spectra were compiled,
red-ward of \lya\ emission and in the rest wavelength of the quasar emission, 
for the two sets of $W_r$ Mg\,II systems described above.
The second and fourth panels of Figure~\ref{fig:dust} display these spectra, 
as well 
as composite `absorption' spectra, for the two Mg\,II absorption $W_r$ bins.  
Also plotted are power-law fits to these spectra, $f(\lambda) \propto 
\lambda^\alpha$.  Dust will
preferentially absorb bluer light, tilting the spectrum and 
lowering the absolute value of $\alpha$.  

The first and third panels represent the result of dividing the composite 
Mg\,II spectra with the control spectra.  
The power-law fits to the 
$1.0$~\AA\ $< W_r < 1.4$~\AA\ composite spectrum and its companion control 
spectrum are nearly identical, $|\Delta(\alpha)| \sim 0.02$.  
These systems, therefore, are unlikely to impose an important dust bias.
The larger $W_r$ sample clearly displays a dust signature,
as seen in the divided spectrum in the upmost panel of 
Figure~\ref{fig:dust}.  For this sample $|\Delta(\alpha)| \sim 0.11$. 
Assuming Small Magellanic Cloud type extinction and employing the 
fitting formula of \cite{pei92}, this $|\Delta(\alpha)|$ corresponds to 
a color excess $E(B-V) \sim 0.01$.  
This is an important result, especially as there are known quasar selection 
effects within SDSS, which uses colors in addition to apparent
magnitude to select quasars for spectroscopic follow-up \citep{richards02}.  
Given the small color excess 
implied by the measured $|\Delta(\alpha)|$, however, 
these effects are likely to small.  
A full consideration of these effects will be the subject of a future
paper.  For now, we note that
the effect of reddening would generally be 
to under-measure \lofx. 
The net effect could be a systematic shallowing of \lofx\, especially 
at $z>1$, although the reddening is sufficiently small that 
the corrections to \lofx\ should be minor.

\subsection{Implications and Interpretations}

The evolution of \lofx\ described by Figure~\ref{fig:dndzSDSS}
has several implications for the physical processes relevant
to \ion{Mg}{2} systems.  First, the near constancy of \lofx\
over the $\approx 3$\,Gyr spanning $z=2$ to 0.8 indicates
the cross-section to strong \ion{Mg}{2} systems is very likely
to be dominated by dark matter halos with masses
$M \ll 10^{12} \msol$.  Although structure forms much
earlier in the $\Lambda$CDM cosmology than an Einstein de Sitter
model, the number density of galaxies with $M > M_*(z=2) \approx 10^{11} \msol$
increases exponentially over this epoch \citep[e.g.][]{jenkins}.
If these massive galaxies dominated the cross-section, then 
$\sigma$ would have to decrease exponentially to yield
constant \lofx.  We consider such a coincidence to be highly
unlikely.  It is far more reasonable to physically associate
the phenomenon with galaxies having mass $M<M_*(z=2)$ and assume
a nearly constant cross-section during this epoch.  Second,
the decline in \lofx\ at $z<0.8$ indicates the cross-section to
strong \ion{Mg}{2} systems is decreasing in time.  This
conclusion applies particularly to the $W_r > 1.8$\AA\ absorbers
where the decrease in \lofx\ is secure.  For the same argument
as above, the number density of dark matter halos is either constant
(for $M \ll M_*$) or increasing substantially (for $M \gg M_*$).
A decrease in \lofx\, therefore, can only be explained by a 
decrease in $\sigma$.  The observations imply the processes
responsible for strong \ion{Mg}{2} absorbers are `turning off' at
$z\sim 1$.  
Third, one identifies a rough correspondence between the 
star formation history of the universe \citep{madau96} and \lofx.
Granted the two quantities track one another over the redshift
range $z=0.5$ to 2, it is tempting to connect the 
strong \ion{Mg}{2} systems with processes related to active
star formation (e.g.\ galactic winds, starbursts).

We now examine the implications of the evolution of \lofx\ in terms
of the two models described in the introduction: (1) a starburst
scenario where the incidence and kinematics 
of strong \ion{Mg}{2} systems are determined by feedback processes
(BCVS);
and (2) infalling gas in the outer halos of massive galaxies (MM96).
An important prediction of the MM96 scenario is that \lofx\ should
steeply decline at high redshift once the characteristic dark matter
halo mass $M_*$ becomes smaller than the typical halo mass giving
rise to strong \ion{Mg}{2} absorption.
In the MM96 scenario, the halo mass is $\sim 10^{12} \msol$,
which is necessary to insure the halo has a significant
reservoir of cool, photoionized gas (MM96).  Within the $\Lambda$CDM cosmology
with $\sigma_8 = 0.9$, $M_* = 10^{12} \msol$ at $z=0.75$.  Therefore,
this model predicts a steep decline in \lofx\ at $z>1$ which is not
observed.
As noted above, to adopt a model where strong \ion{Mg}{2} absorption arises in
an unevolving population of dark matter halos, the typical mass would
need to be less than $M_*(z=2) \approx 10^{10.5} \msol$.
Therefore, we conclude that the majority of strong \ion{Mg}{2}
absorbers do not arise in massive halos typical of the 
$W_r \approx 0.3$\AA\ population at $z\sim 1$.  In fairness, we 
note that the
MM96 model was primarily introduced to explain the $W_r \approx 0.3$\AA\
\ion{Mg}{2} systems which may have a different origin than the strong
\ion{Mg}{2} absorbers considered in this paper. 

It is more difficult to compare the observations against the
starburst scenario because there does not exist a well-developed model.
Nevertheless, consider the following arguments.  At $z \sim 1$, the
dark matter halo merger rate is expected to scale as $(1+z)^3$
\citep{merger}.  If one assumes the starburst phenomenon is merger-driven,
then the simple prediction is an increase in \lofx\ with increasing
redshift.  Indeed, this is observed from $z=0.3$ to 0.8, yet all
of the sub-samples are inconsistent with even a $(1+z)$ increase in
\lofx\ at $z>1$.  
Because the incidence is the product of the number density and
cross-section of the halos, one could envision a scenario where
$\sigma$ is decreasing with redshift as $n$ increases such that
\lofx\ remains roughly constant.  For example, 
the Mg abundance could be much lower at higher redshift or 
the velocity fields more quiescent, 
but these explanations are rather unlikely for a starburst model.
Recent observational results of \cite{deep2} suggest another
possible solution.
These authors find that the galaxy merger rate evolves much more
shallowly with redshift than dark matter haloes, $(1 + z)^{0.51 \pm 0.28}$, 
a much better match to the \ion{Mg}{2} observations.

As noted above, the general trend of \lofx\ roughly
tracks the global density of star formation. 
That is, the star formation rate (SFR) density is described as roughly
constant from $z=1$ to 2 (where it is not well constrained) and 
decreases by a factor of several from $z=1$ to today.
Although the correspondence between the incidence of strong \ion{Mg}{2}
absorbers and the SFR density could be simple coincidence, it
may be expected if the absorption systems arise from feedback processes
related to star formation.  The correspondence may suggest that
significant star formation is required to maintain a large
cross-section to strong \ion{Mg}{2} absorption, i.e.\ this phase is
short-lived in galactic halos and requires continued input to
maintain the observed incidence at $z=1$ to 2.
At the very least, we consider the similarity to be at least
anecdotal evidence in favor of a starburst scenario.

The actual value of the incidence of strong \ion{Mg}{2} absorption 
-- as opposed to its evolution -- provides an important constraint
on the hypotheses of massive galactic halos and starbursts.
The phenomenology 
of MM96 implies a typical impact parameter for \ion{Mg}{2} absorption of 
$\rho \sim 30$~h$^{-1}$~kpc.  Using $\mlofx \sim 3 \times 10^{-5}~
h~{\rm Mpc^{-1}}$ at $z \sim 1$, the value 
measured in this work for $W_r > 1.0$~\AA\ systems,        
a number density $n \sim 0.01 ~h^{3}$~Mpc$^{-3}$ is measured for large, 
quiescent galaxies at this redshift.
It should be recalled that the measured
\lofx\ for \ion{Mg}{2} at this redshift may be somewhat influenced by merging 
systems, so the value used for \lofx\ in this calculation should be 
considered an upper-limit.
Again assuming standard concordance cosmology and using the mass function 
introduced by \cite{jenkins}, the mass of halos with 
$n \sim 0.01 - 0.03 ~h^{3}$~Mpc$^{-3}$ is $M \sim 10^{11}~M_{\odot}$. 
Similar to our previous conclusions, we find that the MM96
scenario is nonviable at $z>1$ if one adopts their typical mass
halo of $10^{12} \msol$ unless one extends the impact parameter to
an unlikely $\rho \sim 100$\,kpc.

Turning to the starburst hypothesis of BCCV, the observations can be used
with previous work on superwinds and LBGs to 
constrain the required impact parameter of the outflows.
If high redshift Mg\,II systems do correspond to SB galaxies, one expects
them to trace LBGs, also thought to harbor starbursts.  Using the measured
number density $n \sim 0.016 ~h^3$~Mpc$^{-3}$ for LBGs at $z \sim 3$
\citep{steidel99}, and assuming the value of \lofx\ observed at $z=2$
for the $W_r > 1$~\AA\ sample ($2 \times 10^{-5}~h~{\rm Mpc^{-1}}$), 
the predicted $\sigma$ for merging Mg\,II systems at $z = 3$ is 
$\sigma \sim 10^{-3} ~h^{-2} ~\textrm{Mpc}^2$.  The value of \lofx\ used in 
this calculation may be considered a maximum, as the overall sample of 
Mg\,II systems discovered in this survey is likely to be polluted with 
large-scale processes, Damped Ly$\alpha$ absorption systems, or other systems,
and thus the resultant value of $\sigma$ is also a maximum.
To determine whether this value is consistent with the wind hypothesis, 
theoretical upper and lower bounds on $\sigma$ may be used.  
Assuming a $v_{bubble} \sim 100 ~\textrm{km} ~\textrm{s}^{-1}$ 
and a bubble age $t_r \sim 60$ Myr (BCCV, a lower limit equal to the 
expected time-scale of the SB), the approximate impact parameter will be 
$\sigma \sim 1 \times 10^{-4} ~\textrm{Mpc}^2$.  
The expansion time and velocity used for this calculation are both 
lower limits, so this is a lower limit.
Computer simulations presented in \cite{FL03} find the maximum size of these bubbles 
(after very long evolutionary periods) to correspond to $\sigma \sim 3 \times 
10^{-2} ~\textrm{Mpc}^{2}$.  
The gas in such large bubbles, however, may be of too low density to give rise
to strong Mg\,II absorption, so this should be considered an extreme upper 
limit.  In any case, the cross-section determined from 
Mg\,II absorption and LBG data is 
consistent with the theory that superwinds are the primary hosts of high 
redshift Mg\,II absorption.  


\section{Kinematics}

Examining the profiles in Figure~\ref{fig:data}, it is evident that the majority
demonstrate at least one of
the three criteria discussed in BCCV for the identification of 
superwind driven galactic bubbles. 
Because these features were clearly present in even the weaker 
$W_r$ systems, the BCCV criterion of 
$W_r > 1.8$~\AA\ has been relaxed to 
$W_r > 1.0 ~\textrm{\AA}$. 
Based primarily on unsaturated transitions (usually a Fe\,II line), the majority 
($\sim$ 80\%) of the discovered systems exhibit some degree of complexity, though 
as discussed above, kinematic features of 
width $\delv \le 45$ \kms are unobservable 
in the ESI spectra.  Because every system discovered in a HIRES
spectrum displays a great deal of spectral complexity, and as many 
of these features are 
kinematically closer than the above specified limit, it is 
probably a safe assumption 
that most or all of the systems studied in this work conform to the complexity 
criterion introduced out by BCCV.  

\begin{figure}
\plotone{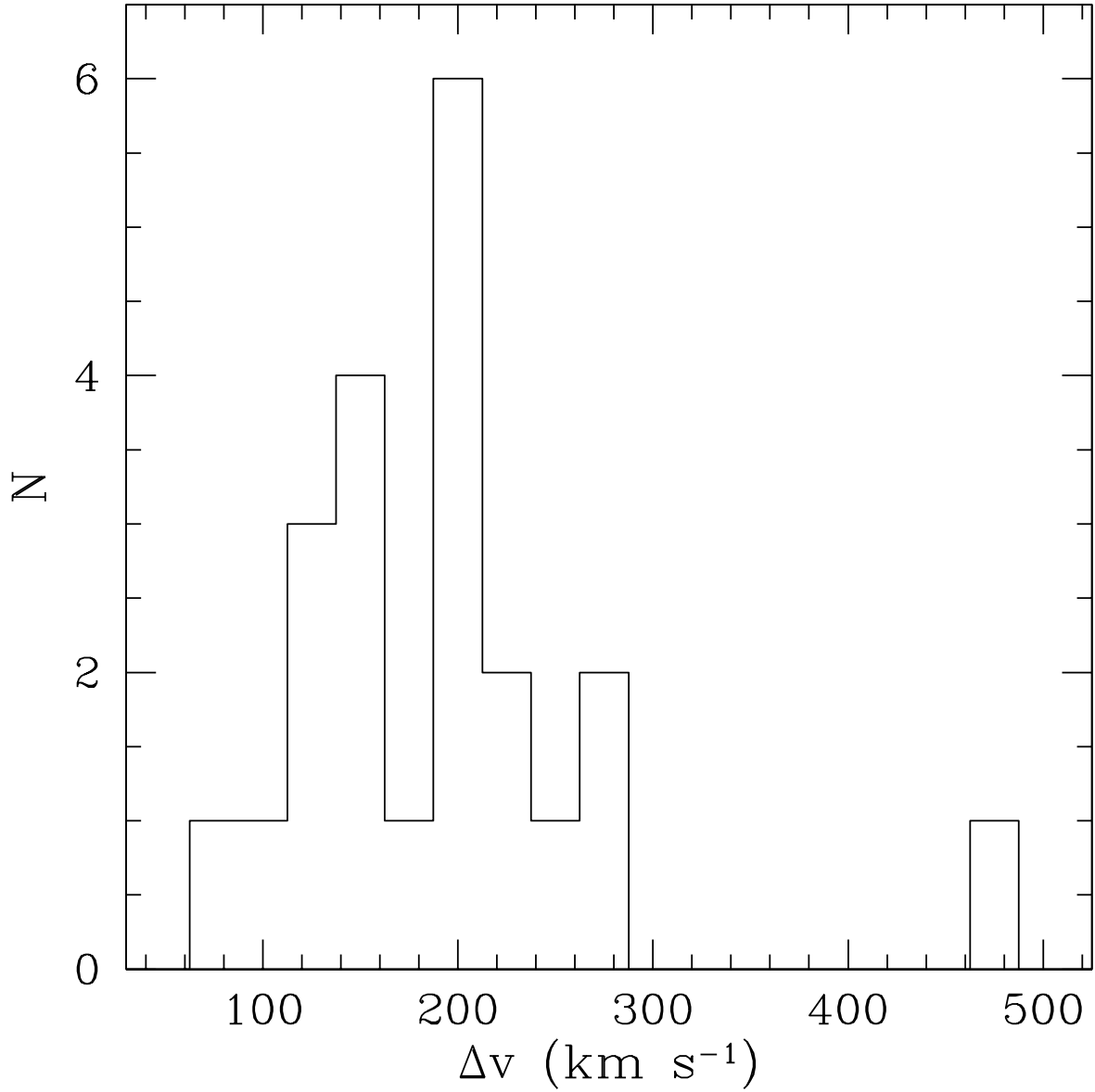}
\caption{A histogram of $\delv$ for the 22 systems discussed in this paper.  Note the 
sharp cutoff at $\sim 100$ \kms corresponds to the $W_r > 1.0$~\AA\ limit imposed 
for consideration of a Mg\,II system.
More importantly, note the near absence of systems with very large $\delv$ value.
\label{fig:delV}}
\end{figure}

Figure~\ref{fig:delV} presents a histogram of the velocity
widths of the high resolution sample.  
Clear is the cutoff at $\sim 100$ \kms which corresponds to the 
$W_r > 1.0$ \AA\ imposed on this study (although recall that the $\delv$
measurements are based on unsaturated profiles). 
The velocities range from 
$\delv = 66 ~\textrm{km} ~\textrm{s}^{-1}$ for the $z=1.328$ system
toward Q1331+17 
to $\delv = 470 ~\textrm{km} ~\textrm{s}^{-1}$ for the $z=1.339$ system
toward
PSS2241+1352, with the majority of systems having $\delv \sim 200 ~\textrm{km} 
~\textrm{s}^{-1}$.  In the idealized case of a spherical shell expanding
with constant velocity $v_{shell}$, the velocity width for an
impact parameter of $R/2$ is $\sqrt{3}v_{shell}$.  Therefore, the median
velocity width for this sample corresponds to a shell speed 
of $v_{shell} \sim 150$~\kms.
This value is consistent, though on the low end, with numbers observed 
for local galactic superwinds \citep{HLSA00} 
and those predicted from numerical simulations 
\citep[e.g., ][]{SS00}.  
An important and somewhat unexpected result of this analysis is that
very few Mg\,II systems show $\delv > 250$ \kms.  If this gas is
tracing winds arising from starburst systems, it is evident that the
optical depth of rapidly expanding 'bubbles' at $z \sim 1.5$ is small: 
$\tau < 1.0 \times 10^{-5}~ h$ Mpc$^{-1}$ (95\% c.l.).

\begin{figure}
\plotone{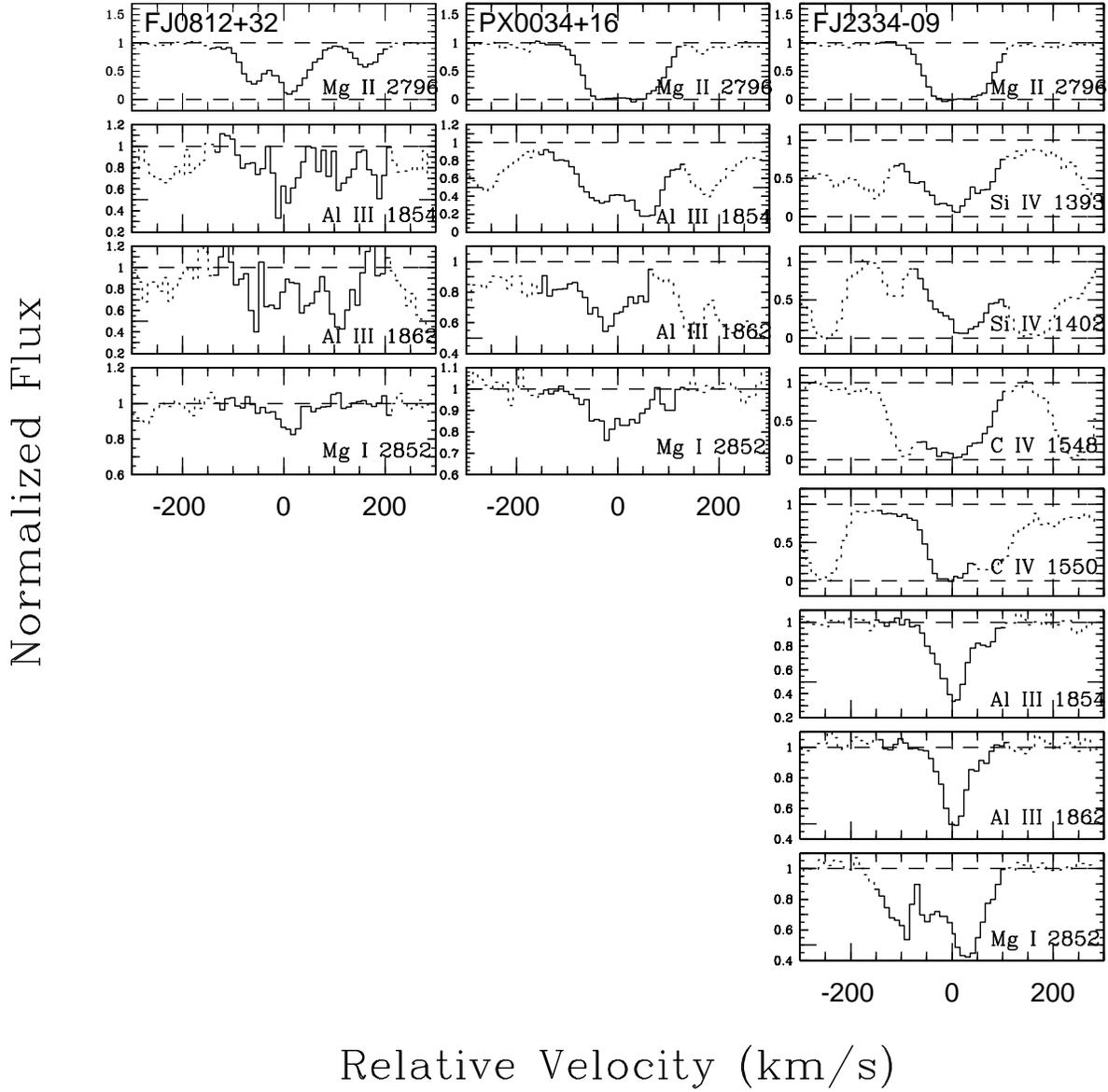}
\caption{Six systems for which transitions of Al\,III, C\,IV, and
Si\,IV are detected.
This figure adopts the same labeling 
as Figure~\ref{fig:data}.
\label{fig:ions}}
\end{figure}

In one respect, the 
data displayed in Figure~\ref{fig:ions} are generally unsupportive 
of the Mg II absorber-superwind hypothesis.  
In a wind model, one envisions a cold, expanding shell of gas which 
envelopes a rarefied, hot region which in turn 
surrounds the post-SB galaxy.   In this scenario, one would 
not expect to observe high-ionization state particles 
(e.g. Si IV) in the same distribution 
as the colder gas (Mg II; \cite{SS00}).  
One might expect the higher-ionization state 
gas to be distributed in a kinematically narrower region (it is no longer 
expanding).  Alternatively, the gas may exhibit turbulent velocities at
the virial velocity dispersion of the galaxy
and not conform to the clump patterns of the gas in the shell.  In
either case, these 
trends are not observed in those systems with high-ionization lines.  
Figure~\ref{fig:ions} shows a clear trend for high- and low-ionization state gas
to share the same regions of \kms space and to often 
display similar clumping patterns.  
This result is consistent with the work of \cite{CEA99}, who 
reported strong kinematic links between Mg\,II and C\,IV absorption in 
the redshift range $0.4 \leq z \leq 1.4$, and discussed consequences for both 
the above scenario and that proposed by MM96, which
involves cold bubbles of gas confined by pressure in 
a hot gas medium.  

The construction of MM96 also implies different kinematic situations for 
the hot and cold gas, a prediction not supported by the data.  Indeed, the 
models presented in MM96 predict that systems showing strong Mg\,II 
absorption (large galaxies) should contain little if any C\,IV absorption, 
which is clearly ruled out by the data.
A suggestion put forward to fix this discrepancy is a "warm"
phase of collisionally ionized gas in an interface between the cold bubbles and 
the hot medium.  This would explain the observed kinematic link between Mg\,II 
and more highly ionized species (in either the expanding bubble or 
two-phase gaseous halo scenarios), though the authors point out that the 
amount of collisionally ionized C\,IV found in such a phase would likely 
be small.  There is also the possibility that C\,IV
arises at higher radii than the simple models of MM96 consider.  
This solution, while providing for more C\,IV absorption,  does not provide a simple 
explanation for the observed kinematic link between species.  More sophisticated models of 
the two-phase halo structure are needed to clarify this issue.

\section{Concluding Remarks}


There are many open questions left in the task of assigning a 
particular phenomenology to strong \ion{Mg}{2} absorption.
Detailed simulations of superwinds would be useful in pinning down the 
expected absorption profile of such a system, as well as providing information 
on expected metallicities and temperatures.
A handful of strong Mg II absorption-selected galaxies have been studied
kinematically \citep[see][]{steidel02,CMC00}, 
with varying results, though the number remains small and little consideration 
has gone into the link stipulated in this work.  A more comprehensive survey 
over a larger redshift range would allow for stronger statements regarding the 
correlation between these absorbers and galaxy populations, as well as a 
refined understanding of their kinematic nature.  Optical studies could only 
provide results at low redshift,  but confirmation of the SB and post-SB 
nature expected of the hosting galaxies, even for low-z examples, would 
strengthen the argument considerably.  Adaptive optics might be useful 
here, and a program has been begun at Lick Observatory to obtain AO 
images of a few quasars known to harbor low redshift, strong \ion{Mg}{2} 
systems
in an attempt to morphologically identify these systems.  
We are also performing a separate imaging program with ACS on 
the Hubble Space Telescope (Cycle 14) to cover a wider range of 
redshift.
Finally, with the advent of infrared spectrographs
(e.g. the GNIRS instrument on Gemini South) and the availability of large 
numbers of bright, high redshift quasars (as in the SDSS DR3),
the phenomenon can be studied 
to very high redshift ($z \sim 4$). 
In particular, it is critical to further test 
the assertion that strong \ion{Mg}{2} systems 
trace processes related to star formation.

\acknowledgments
The authors wish to recognize and acknowledge the very significant cultural
role and reverence that the summit of Mauna Kea has always had within the
indigenous Hawaiian community.  We are most fortunate to have the
opportunity to conduct observations from this mountain.
We also acknowledge the Keck support staff for their efforts
in performing these observations.  
We acknowledge the tremendous effort put forth by the SDSS team to
produce and release the SDSS survey.
We thank M. Kuhlen, J. Primack, and T.J. Cox for helpful discussions.
Finally, we would like to thank the anonymous referee of this paper who 
generated a number of useful comments and suggestions. 
GEP and JXP are supported by NSF grant AST-0307408.

\begin{figure*}
\begin{center}
\plotone{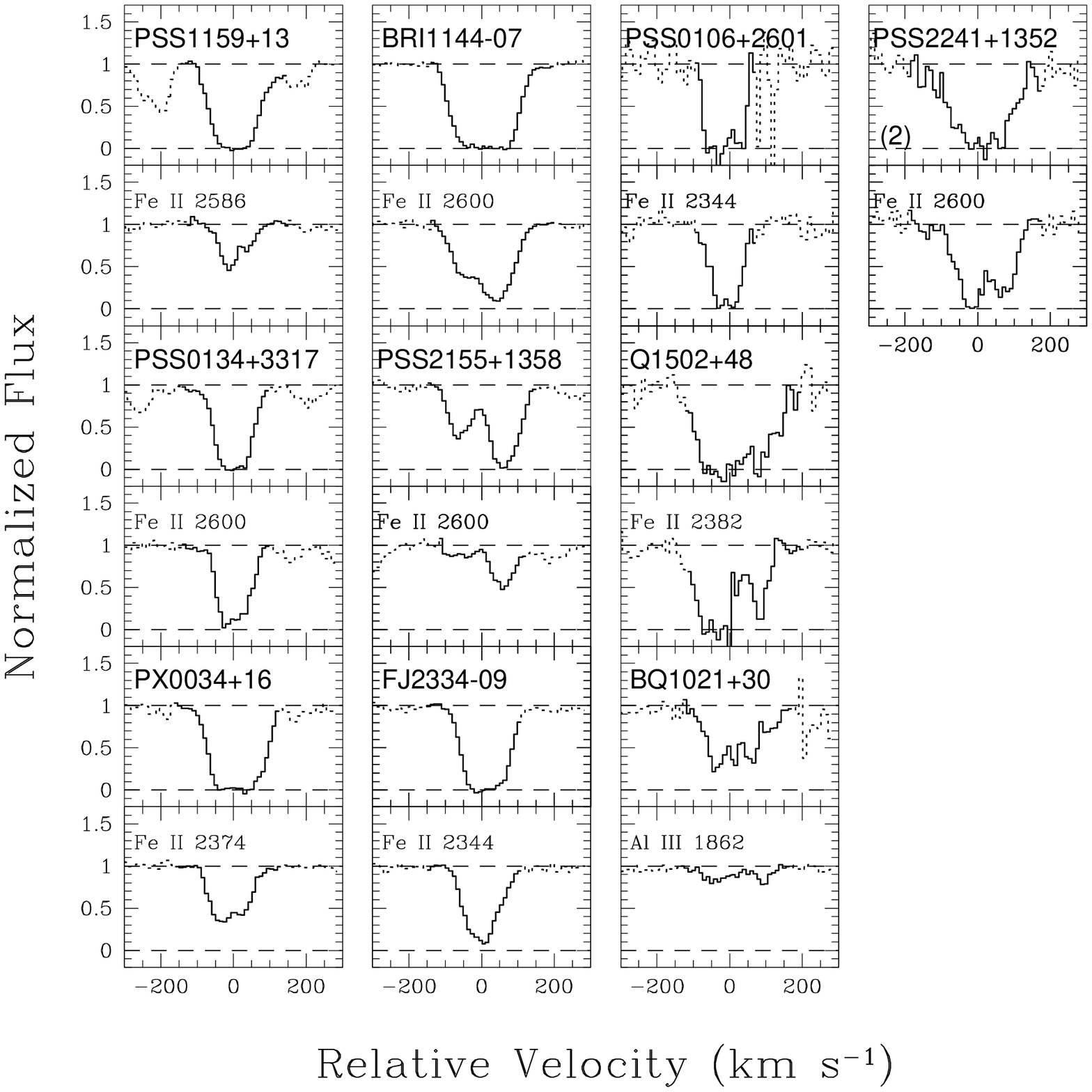}
\end{center}
\end{figure*}

\begin{figure*}
\begin{center}
\plotone{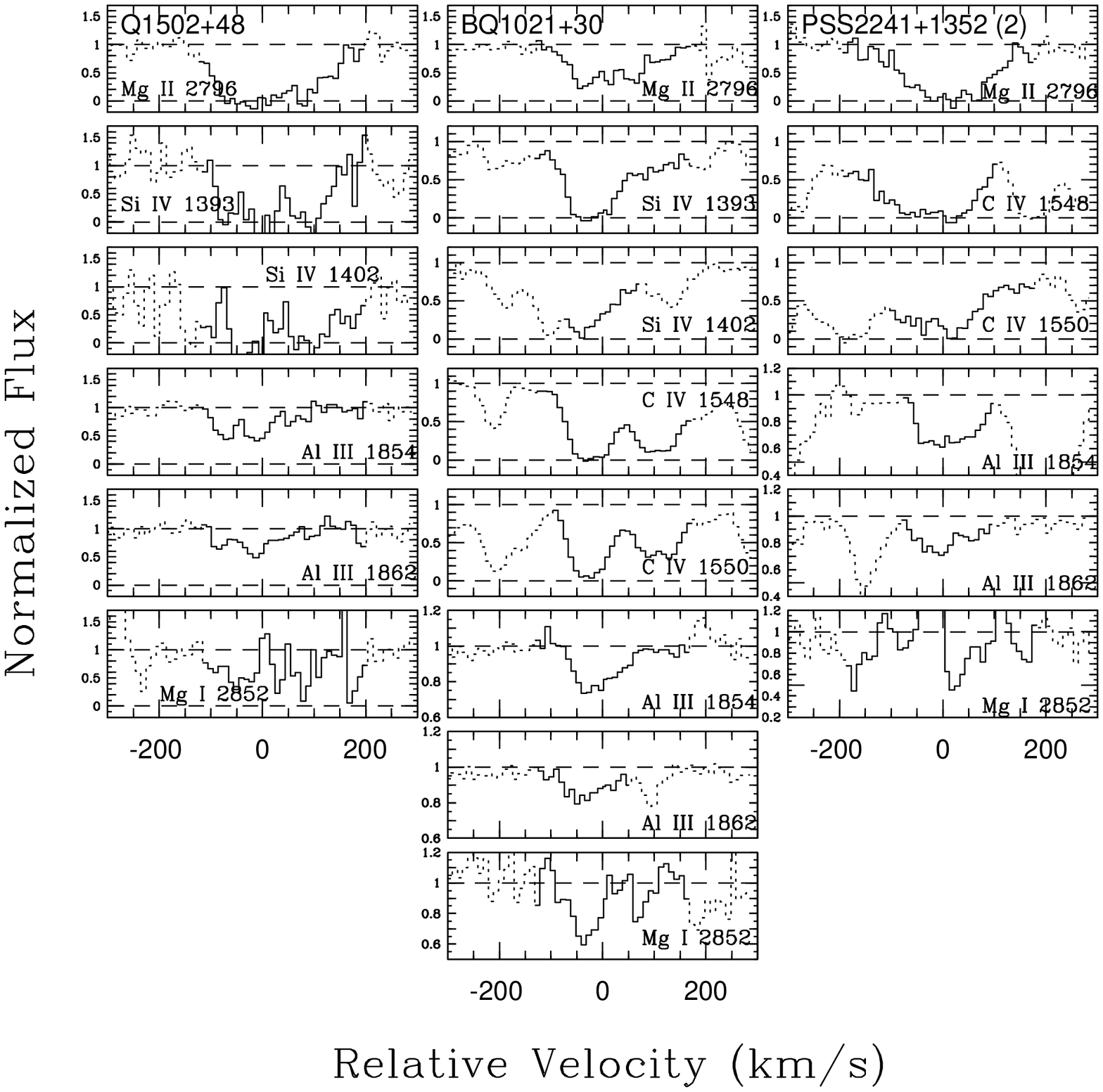}
\end{center}
\end{figure*}

\clearpage
\begin{deluxetable}{cccccc}
\tablecolumns{10}
\tablewidth{0pc}
\tablecaption{SDSS-DR3 $W_r(2796) > 1.0$~\AA\ Mg\,II Systems}
\tablehead{
\colhead{Quasar} & \colhead{RA} & \colhead{Dec} & \colhead{$z_{qso}$} & \colhead{$z_{Mg\,II}$} & \colhead{$W_r(2796)$}}
\startdata
J000009.42-102751.87& 00:00:09.42 &-10:27:51.87&1.844&1.313&1.69\\
J000050.60-102155.87& 00:00:50.60 &-10:21:55.87&2.640&1.056&1.17\\
J000053.72+150637.20& 00:00:53.72 &+15:06:37.20&1.411&0.897&1.63\\
J000143.41+152021.49& 00:01:43.41 &+15:20:21.49&2.638&1.598&1.03\\
J000221.11+002149.46& 00:02:21.11 &+00:21:49.46&3.057&1.958&2.33\\
J000221.80+151454.58& 00:02:21.80 &+15:14:54.58&1.823&1.172&1.05\\
J000221.80+151454.58& 00:02:21.80 &+15:14:54.58&1.823&1.435&1.29\\
J000221.80+151454.58& 00:02:21.80 &+15:14:54.58&1.823&0.716&1.22\\
J000221.80+151454.58& 00:02:21.80 &+15:14:54.58&1.823&0.846&1.68\\
J000341.22-094101.03& 00:03:41.22 &-09:41:01.03&1.739&1.584&2.39\\
\enddata
\tablecomments{The complete version of this table is in the electronic 
edition of the Journal.  The printed edition contains only a sample.} 
\label{tab:sdss}
\end{deluxetable}

\clearpage
\begin{deluxetable}{cccccccccc}
\tablecolumns{10}
\tablewidth{0pc}
\tablecaption{High Resolution Survey Quasars and Found Mg II Systems}
\tablehead{
\colhead{QSO} & \colhead{Inst} & \colhead{$z_{qso}$} & \colhead{$z_{min}$} & \colhead{$z_{max}$} &
\colhead{$z_{Mg\,II}$} & \colhead{$\Delta v$} & \colhead{$W_r$} & \colhead{Metals} & 
\colhead{components}}
\startdata
BR0019-15 & 1 & 4.53 & 1.40 & 1.98&&&&& \\
BR0951-04 & 1 & 4.37 & 1.36 & 2.00&&&&& \\
HS0741+47 & 1 & 3.22 & 0.81 & 1.67&&&&& \\
J0255+00 & 1 & 3.97 & 1.16 & 1.92&&&&& \\
PH957 & 1,2 & 2.69 & 0.47 & 2.76&&&&& \\
PSS0957+33 & 1,2 & 4.25 & 1.29 & 2.65&&&&& \\
PSS1443+27 & 1 & 4.41 & 1.38 & 2.12&&&&& \\
Q0000-26 & 1 & 4.11 & 1.26 & 1.91&&&&& \\
Q0149+33 & 1 & 2.43 & 0.46 & 1.33&&&&& \\
Q0201+11 & 1 & 3.61 & 1.01 & 1.73&&&&& \\
Q0201+36 & 1 & 2.49 & 0.69 & 1.57&&&&& \\
Q0249-22 & 1 & 3.20 & 0.30 & 0.80&&&&& \\
Q0322-32 & 1& & 0.42 & 0.92&&&&& \\
Q0336-01 & 1 & 3.20 & 0.80 & 1.28&&&&& \\
Q0347-38 & 1 & 3.23 & 0.84 & 1.62 & 1.457 & 216 & 2.237& & 7 \\
Q0450-13 & 1 & 2.30 & 0.34 & 0.76&&&&& \\
Q0458-02 & 1 & 2.29 & 0.41 & 1.25&&&&& \\
Q0551-36 & 1 & 2.32 & 0.42 & 1.23&&&&& \\
Q0757+52 & 1& & 0.32 & 0.83&&&&& \\
Q0836+11 & 1 & 2.70 & 0.52 & 1.57&&&&& \\
Q0841+12 & 1 & 2.20 & 0.71 & 1.57 & 1.098 & 208 & 1.450& & 4 \\
Q0841+12&&&& & 1.219 & 274 & 2.388& & 5 \\
Q0930+28 & 1 & 3.42 & 0.86 & 1.56&&&&& \\
Q0952-01 & 1 & 4.43 & 1.36 & 1.91&&&&& \\
Q1005+36 & 1& & 0.80 & 1.51&&&&& \\
Q1055+46 & 1 & 4.13 & 1.25 & 1.86 & 1.386 & 210 & 1.164& & 4 \\
Q1104-18 & 1 & 2.31 & 0.44 & 1.47&&&&& \\
Q1108-07 & 1 & 3.92 & 1.12 & 1.98&&&&& \\
Q1202-07 & 1 & 4.69 & 1.49 & 2.00&&&&& \\
Q1210+17 & 1 & 2.54 & 0.56 & 1.21&&&&& \\
Q1215+33 & 1 & 2.61 & 0.39 & 1.26&&&&& \\
Q1223+17 & 1 & 2.92 & 0.71 & 1.57&&&&& \\
Q1331+17 & 1 & 2.08 & 0.51 & 1.37 & 1.328 & 66 & 1.189& & 2 \\
Q1346-03 & 1 & 3.99 & 1.18 & 1.98&&&&& \\
Q1425+60 & 1 & 3.17 & 0.81 & 1.21&&&&& \\
Q1759+75 & 1 & 3.05 & 1.14 & 1.71&&&&& \\
Q1850+40 & 1 & 2.12 & 0.82 & 1.79&&&&& \\
Q2038-01 & 1 & 2.78 & 0.65 & 1.21&&&&& \\
Q2206-19 & 1 & 2.56 & 0.52 & 1.33&&&&& \\
Q2223+20 & 1,2 & 3.56 & 0.99 & 2.65&&&&& \\
Q2230+02 & 1 & 2.51 & 0.38 & 1.46 & 1.059 & 110 & 1.234& & 4 \\
Q2231-00 & 1 & 3.02 & 0.74 & 1.29&&&&& \\
Q2233+13 & 1 & 3.27 & 0.86 & 1.61&&&&& \\
Q2343+12 & 1 & 2.52 & 0.43 & 1.28&&&&& \\
Q2344+12 & 1 & 2.79 & 0.55 & 1.17&&&&& \\
Q2348-01 & 1 & 3.01 & 0.81 & 1.68&&&&& \\
Q2358-02 & 1& & 0.66 & 1.33&&&&& \\
BQ1021+30 & 2 & 3.12 & 0.73 & 2.65 & 2.329 & 200 & 1.023 & SiIV,CIV,AlIII & 3 \\
BRI1144-07 & 2 & 4.16 & 1.26 & 2.65 & 1.908 & 190 & 1.833& & 2 \\
BRJ0426-2202 & 2 & 4.30 & 1.32 & 2.65&&&&& \\
CTQ460 & 2 & 3.13 & 0.68 & 2.65&&&&& \\
FJ0747+2739 & 2 & 4.11 & 1.25 & 2.65&&&&& \\
FJ0812+32 & 2 & 2.70 & 0.63 & 2.65 & 1.223 & 260 & 1.200& & 3 \\
FJ0812+32&&&& & 2.626 & 280 & 1.300 & SiIV,CIV,AlIII & 1 \\
FJ2129+00 & 2 & 2.94 & 0.72 & 2.65&&&&& \\
FJ2334-09 & 2 & 3.33 & 0.90 & 2.65 & 2.152 & 140 & 1.363 & SiIV,CIV,AlIII & 1 \\
HS1132+22 & 2 & 2.88 & 0.66 & 2.65&&&&& \\
HS1437+30 & 2 & 2.99 & 0.70 & 2.65&&&&& \\
PC0953+47 & 2 & 4.46 & 1.40 & 2.50&&&&& \\
PKS1354-17 & 2 & 3.15 & 0.79 & 2.65&&&&& \\
PSS0106+2601 & 2 & 4.32 & 1.32 & 2.65 & 2.198 & 130 & 1.110& & 1 \\
PSS0133+0400 & 2 & 4.13 & 1.25 & 2.65 & 1.666 & 160 & 1.632& & 2 \\
PSS0134+3317 & 2 & 4.52 & 1.40 & 2.65 & 1.757 & 130 & 1.145& & 1 \\
PSS0209+0517 & 2 & 4.18 & 1.26 & 2.65 & 1.282 & 380 & 1.756& & 1 \\
PSS0808+52 & 2 & 4.45 & 1.40 & 2.65&&&&& \\
PSS1159+13 & 2 & 4.07 & 1.22 & 2.65 & 1.739 & 130 & 1.386& & 2 \\
PSS1248+31 & 2 & 4.35 & 1.32 & 2.65&&&&& \\
PSS1253-02 & 2 & 4.01 & 1.18 & 2.65&&&&& \\
PSS1315+29 & 2 & 4.18 & 1.29 & 2.65&&&&& \\
PSS1432+39 & 2 & 4.28 & 1.31 & 2.65&&&&& \\
PSS1506+52 & 2 & 4.18 & 1.25 & 2.65 & 1.472 & 280 & 3.407& & 2 \\
PSS1723+2243 & 2 & 4.52 & 1.43 & 2.65&&&&& \\
PSS2155+1358 & 2 & 4.26 & 1.29 & 2.65 & 1.914 & 200 & 1.287& & 2 \\
PSS2241+1352 & 2 & 4.44 & 1.32 & 2.65 & 1.339 & 470 & 4.626& & 2 \\
PSS2241+1352&&&& & 2.542 & 170 & 1.772& & 2 \\
PSS2323+2758 & 2 & 4.18 & 1.25 & 2.65&&&&& \\
PSS2344+0342 & 2 & 4.30 & 1.29 & 2.65&&&&& \\
PX0034+16 & 2 & 4.29 & 1.32 & 2.65 & 1.800 & 160 & 1.576& & 2 \\
Q0112-30 & 2 & 2.96 & 0.71 & 2.65&&&&& \\
Q0821+31 & 2 & 2.61 & 0.61 & 2.65 & 2.535 & 190 & 1.738 & SiIV,CIV,AlIII & 2 \\
Q0933+28 & 2& & 0.93 & 2.65&&&&& \\
Q1209+09 & 2 & 3.30 & 0.86 & 2.65 & 2.585 & 270 & 3.020 & SiIV,CIV,AlIII & 2 \\
Q1337+11 & 2 & 2.92 & 0.68 & 2.65&&&&& \\
Q1502+48 & 2 & 3.20 & 0.82 & 2.65 & 2.272 & 220 & 1.807 & AlIII & 2 \\
Q2342+34 & 2 & 3.01 & 0.77 & 2.65&&&&& \\
SDSS0127-00 & 2 & 4.06 & 1.22 & 2.65&&&&& \\
SDSS2350-00 & 2 & 3.01 & 0.75 & 2.65 & 0.863 & 200 & 1.474& & 3 \\
SDSS2350-00&&&& & 2.425 & 280 & 2.732 & SiIV,CIV,AlIII & 2 \\
\enddata
\label{tab:high}
\end{deluxetable}

\clearpage
\begin{deluxetable}{ccccccc}
\tablecolumns{11}
\tablewidth{0pc}
\tablecaption{SDSS-DR3 Mg\,II $dN/dX$}
\tablehead{
\colhead{$<z>$} & \colhead{$dN/dX$} & \colhead{N} &
\colhead{$dN/dX$} & \colhead{N} &
\colhead{$dN/dX$} & \colhead{N} \\
&\multicolumn{2}{c}{$W_r > 1.0$~\AA}&
\multicolumn{2}{c}{$1.0$~\AA\ $< W_r < 1.4$~\AA}&
\multicolumn{2}{c}{$W_r > 1.8$~\AA}}
\startdata
0.41 & $0.056 \pm 0.006$ & 96 & $0.030 \pm 0.004$ & 52 & $0.009 \pm 0.002$ & 16 \\
0.50 & $0.059 \pm 0.004$ & 190 & $0.030 \pm 0.003$ & 97 & $0.012 \pm 0.002$ & 41 \\
0.60 & $0.066 \pm 0.004$ & 253 & $0.033 \pm 0.003$ & 128 & $0.016 \pm 0.002$ & 61 \\
0.70 & $0.081 \pm 0.004$ & 351 & $0.036 \pm 0.003$ & 157 & $0.025 \pm 0.002$ & 109 \\
0.79 & $0.088 \pm 0.004$ & 403 & $0.045 \pm 0.003$ & 206 & $0.021 \pm 0.002$ & 96 \\
0.89 & $0.089 \pm 0.004$ & 417 & $0.045 \pm 0.003$ & 211 & $0.023 \pm 0.002$ & 106 \\
0.98 & $0.080 \pm 0.004$ & 328 & $0.036 \pm 0.003$ & 147 & $0.022 \pm 0.002$ & 91 \\
1.08 & $0.091 \pm 0.005$ & 297 & $0.038 \pm 0.003$ & 124 & $0.028 \pm 0.003$ & 91 \\
1.18 & $0.092 \pm 0.005$ & 396 & $0.043 \pm 0.003$ & 187 & $0.026 \pm 0.002$ & 112 \\
1.27 & $0.089 \pm 0.004$ & 388 & $0.043 \pm 0.003$ & 185 & $0.023 \pm 0.002$ & 101 \\
1.37 & $0.102 \pm 0.005$ & 451 & $0.048 \pm 0.003$ & 215 & $0.026 \pm 0.002$ & 116 \\
1.46 & $0.088 \pm 0.005$ & 336 & $0.039 \pm 0.003$ & 149 & $0.023 \pm 0.002$ & 89 \\
1.56 & $0.097 \pm 0.005$ & 315 & $0.040 \pm 0.003$ & 130 & $0.034 \pm 0.003$ & 110 \\
1.66 & $0.100 \pm 0.006$ & 249 & $0.044 \pm 0.004$ & 108 & $0.030 \pm 0.003$ & 74 \\
1.75 & $0.084 \pm 0.007$ & 129 & $0.036 \pm 0.005$ & 55 & $0.023 \pm 0.004$ & 35 \\
1.85 & $0.089 \pm 0.009$ & 100 & $0.039 \pm 0.006$ & 44 & $0.027 \pm 0.005$ & 30 \\
1.94 & $0.094 \pm 0.010$ & 84 & $0.036 \pm 0.006$ & 32 & $0.035 \pm 0.006$ & 31 \\ 
2.04 & $0.076 \pm 0.013$ & 33 & $0.016 \pm 0.006$ & 7 & $0.032 \pm 0.009$ & 14 \\
2.14 & $0.070 \pm 0.020$ & 12 & $0.012 \pm 0.008$ & 2  & $0.029 \pm 0.013$ & 5 \\ 
2.23 & $0.065 \pm 0.024$ & 7 & $0.046 \pm 0.021$ & 5 & $0.009 \pm 0.009$ & 1 \\ 
\enddata
\tablecomments{All $dN/dX$ values reported are multiplied by $c/H_0$ for 
ease of readability, and are in units of $h~Mpc^{-1}$.} 
\label{tab:dndz}
\end{deluxetable}

\clearpage
\begin{deluxetable}{ccc}
\tablecolumns{10}
\tablewidth{0pc}
\tablecaption{$dN/dX$ and $dN/dz$ Functional Fits to SDSS-DR3 \ion{Mg}{2} Systems}
\tablehead{
\colhead{$W_r$(\AA)} & \colhead{Shape$^a$} & \colhead{Normalization}}
\startdata
\multicolumn{3}{c}{$dN/dX = N + m z$} \\
      $>1.0$ & $0.022 \pm 0.004$ & $0.036 \pm 0.004$ \\
      $>1.8$ & $0.011 \pm 0.002$ & $0.011 \pm 0.002$ \\
$ 1.0 < W_r < 1.4$ & $0.002 \pm 0.003$ & $0.036 \pm 0.002$ \\
\hline
\multicolumn{3}{c}{$dN/dX = N \exp^{-z_0/z}$} \\
      $>1.0$ & $0.275 \pm 0.05$ & $0.113 \pm 0.006$ \\
      $>1.8$ & $0.467 \pm 0.10$ & $0.037 \pm 0.003$ \\
$ 1.0 < W_r < 1.4$ & $0.123 \pm 0.07$ & $0.044 \pm 0.003$ \\
\hline
\multicolumn{3}{c}{$dN/dX = N(1+z)^{\gamma}$} \\
      $>1.0$ & $0.54 \pm 0.10$ & $0.06 \pm 0.004$ \\
      $>1.8$ & $0.98 \pm 0.154$ & $0.01 \pm 0.001$ \\
$ 1.0 < W_r < 1.4$ & $0.13 \pm 0.15$ & $0.04 \pm 0.004$ \\
\hline
\multicolumn{3}{c}{$dN/dz = N(1+z)^{\gamma}$} \\
      $>1.0$ & $1.40 (1.24,1.56)$ & $0.08 (0.095,0.075)$ \\
      $>1.4$ & $1.74 (1.52,1.96)$ & $0.036 (0.042,0.030)$ \\
      $>1.8$ & $1.92 (1.60,2.22)$ & $0.016 (0.021,0.013)$ \\
$ 1.0 < W_r < 1.4$ & $0.99 (0.77,1.28)$ & $0.051 (0.060,0.041)$ \\
$ 1.4 < W_r < 1.8$ & $1.56 (1.25,1.89)$ & $0.020 (0.025,0.015)$ \\
\enddata
\tablenotemark{a}{$m$, $z_0$, or $\gamma$}
\tablecomments{These are the $1 \sigma$ errors for the 
minimized-$\chi^2$
power-law fits and the 95\% Confidence Limits for the maximum likelihood 
power-law fits to the SDSS-DR3 $dN/dz$ data.  For the $dN/dz$ results, the 
upper and lower Confidence Limits for $\gamma$ and $N$ are presented in 
parentheses.  The values of $N$ for these fits are 
calculated by constraining the integral of $g(z)*dN/dz$ to equal the 
observed number of systems for a given cut in $W_r$.}
\label{tab:errors}
\end{deluxetable}


\begin{thebibliography}{}
\bibitem[Adelberger et al.(2003)]{ASSP03} Adelberger, K. L., Steidel, C. C., Shapley, 
     A. E., \& Pettini, M. 2003, \apj, 584, 45

\bibitem[Bolton et al.\ (2004)]{BBS04} Bolton, A. S., Burles, S., Schlegel, D. J., 
     Eisenstein, D. J., \& Brinkmann, J. 2004, \aj, 127, 1860

\bibitem[Bond et al.\  (2001a)]{BCCV} Bond, N. A., Churchill, C. W., Charlton, 
     J. C, \& Vogt, S. S. 2001, \apj, 562, 641

\bibitem[Bond et al.\ (2001b)]{bccv2} Bond, N. A., Churchill, C. W., Charlton, 
     J. C, \& Vogt, S. S. 2001, \apj, 557, 761

\bibitem[Chen et. al.\ (2003)]{chen03} Chen, H. W., Marzke, R. O., McCarthy, 
     P. J., Martini, P., Carlberg, R. G., Persson, S. E., Bunker, A., Bridge, 
     C. R., \& Abraham, R. G. 2003, \apj, 586, 745

\bibitem[Churchill et al.\ (2004)]{CKS04} Churchill, C. W., Kacprzak, G. G., 
     Steidel, C. C. AAS Meeting 205, 129.04

\bibitem[Churchill et al.\ (2000)]{CMC00} Churchill, C. W., Mellon, R. R., 
     Charlton, J. C., Jannuzi, B. T., Kirhakos, S., Steidel, C. C., \&
     Schneider, D. P. 2000, \apj, 543, 577

\bibitem[Churchill et al.\ (1999)]{CEA99} Churchill, C. W., Mellon, R. R., 
     Charlton, J. C., Jannuzi, B. T., Kirhakos, S., Steidel, C. C., \& 
     Schneider, D. P. 1999, \apj, 519, L43 

\bibitem[Cox et al.\ (2004)]{cox04} Cox, T. J., Primack, J., Jonsson, P., \&
     Somerville, R. 2004, preprint (astro-ph:0402675)

\bibitem[Furlanetto \& Loeb \ (2003)]{FL03} Furlanetto, S. R. \& Loeb, A. 2003, 
     \apj, 588, 18

\bibitem[Heckman et al.\ (2000)]{HLSA00} Heckman, T. M., Lehnert, M. D., Strickland, 
     D. K., \& Armus, L. 2000, \apjs, 129, 493

\bibitem[Heckman et al.\ (2001)]{HSM01} Heckman, T. M., Sembach, K. R., Meurer, G. R., 
     Strickland, D. K., Martin, C. L., Calzetti, D., Leitherer, C. 2001, /apj, 554, 1021

\bibitem[Holmberg \ (1975)]{hberg75} Holmberg, E. 1975, in {\it Stars and Stellar
     Systems, 9, Galaxies and the Universe}, ed(s)., A. Sandage, M. Sandage, 
     \& J. Kristian, (Chicago: Univ. Chicago Press), p. 123

\bibitem[Jenkins et al.\  (2001)]{jenkins} Jenkins, A., Frenk, C. S., White, D. M.,
     Colberg, J. M., Cole, S., Evrard, A. E., Couchman, H. M. P., \& Yoshida, N. 
     2001, \mnras, 321, 372

\bibitem[Lanzetta, Wolfe, \& Turnshek\  (1987)]{lwt87}		
Lanzetta, K. M., Wolfe, A. M., \&  Turnshek 1987, \apj, 322, 739

\bibitem[Lanzetta\  (1993)]{lzt93}
Lanzetta, K.M. 1993,  in {\it The Environment and Evolution
of Galaxies}, ed. J.M. Shull \& H.A. Thronson, Jr.
(Boston: Kluwer Academic Publishers), p. 237

\bibitem[Lanzetta et al.\ (2002)]{lzt02}  Lanzetta, K. M., Yahata, N., 
     Pascarelle, S., Chen, H. W., Fernandez-Soto, A. 2002, \apj, 570, 492

\bibitem[Len et al.\ (2004)]{deep2} Lin, L., Koo, D. C., Willmer, C. N. A., 
     Patton, D. R., Conselice, C. J., Yan, R., Coil, A. L., Cooper, M. C., 
     Davis, M., Faber, S. M., Gerke, B. F., Guhathakurta, P., Newman, J. A. 2004, 
     \apjl, 617, L9

\bibitem[Madau et al.\ (1996)]{madau96}		
Madau, P., Ferguson, H.C., Dickinson, M.E., Giavalisco, M., 
Steidel, C.C., \& Fruchter, A. 1996, \mnras, 283, 1388

\bibitem[Madau et al.\ (2001)]{MFR} Madau, P., Ferrara, A., \& Rees, M. J. 2001,
     \apj, 555, 92

\bibitem[Martin et al.\ (2002)]{MKH} Martin, C. L., Kobulnicky, H. A., \&
     Heckman, T. M. 2002, \apj, 574, 663

\bibitem[Menard \ (2005)]{menard05} Menard, B. 2005, \apj, 630, 28

\bibitem[Misawa et al.\ (2002)]{dndz} Misawa, T., Tytler, D., Iye, M., 
     Storrie-Lombardi, L. J., Suzuki, N., \& Wolfe, A. M. 2002, \aj, 123, 1847

\bibitem[Mo \& Miralda-Escude \ (1996)]{MM96} Mo, H. J. \& Miralda-Escude, J. 1996,
     \apj, 469, 589

\bibitem[Murphy \& Liske\  (2004)]{murphy04} Murphy, M. T., Liske, J. 2004,
     \mnras, 354, 31

\bibitem[Nestor et al.\ (2003)]{nestor03} Nestor, D. B., Rao, S., M., 
     Turnshek, D. A., \& Vanden Berk, D. 2003, /apj, 595, 5

\bibitem[Nestor et al.\ (2004)]{sdssedr} Nestor, D. B., Turnshek, D. A., 
     \& Rao, S. M. 2004, astro-ph/0410493

\bibitem[Patton et al.\ (1997)]{merger} Patton, D. R., Pritchet, C. J., Yee, H. K., 
     Ellingson, E., \& Carlberg, R. G. 1997, \apj, 475, 29

\bibitem[Pei et al.\ (1991)]{pei91} Pei, Y. C., Fall, S. M., Bechtold, J. 
     1991, \apj, 378, 6

\bibitem[Pei (1992)]{pei92} Pei, Y. C. 1992, \apj, 395, 130

\bibitem[Pettini et al.\ (1998)]{PEA98} Pettini, M., Kellogg, M., Steidel, C. C., 
     Dickenson, M., Adelberger, K. L., \& Giavalisco, M. 1998, \apj, 508, 539

\bibitem[Pettini et al.\ (2001)]{pettini01} Pettini, M., Shapley, A. E., Steidel, C. C.,
     Cuby, J-G., Dickinson, M., Moorwood, A. F. M., Adelberger, K. L., Giavalisco, M.
     2001, \apj, 554, 981

\bibitem[Pettini et al.\ (2003)]{pettini03}  	
Pettini, M., Madau, P., Bolte, M., Prochaska, J.X.,
Ellison, S.L., \& Fan, X. 2003, \apj, 594, 695

\bibitem[Prochaska and Wolfe\ (1997)]{PP97} Prochaska, J. X., \& Wolfe, M. W. 
     1997, \apj, 487, 73


\bibitem[Prochaska et al.\ (2001)]{PWT01} Prochaska, J. X., Wolfe, A. M., 
     Tytler, D., Burles, S., Cooke, J., Gawiser, E., Kirkman, D., O'Meara, 
     J. M., \& Storrie-Lombardi, L. 2001, \apjs, 137, 21

\bibitem[Prochaska et al.\ (2003)]{pro03}  		
Prochaska, J.X., Gawiser, E., Wolfe, A.M., Cooke, J.,
\& Gelino, D. 2003, \apjs, 147, 227


\bibitem[Richards et al.\ (2002)]{richards02}
Richards, G.T. et al.\ 2002, \aj, 123, 2945

\bibitem[Scannapieco et al.\ (2002)]{SFM} Scannapieco, E., Ferrara, A., \&
     Madau, P. 2002, \apj, 574, 590

\bibitem[Schaye\  (2001)]{Schaye} Schaye, J. 2001, \apjl, 559, L1

\bibitem[Schaye et al.\ (2003)]{schaye03}       
Schaye, J., Aguirre, A., Kim, T.-S., Theuns, T., Rauch, M., 
\& Sargent, W.L.W. 2003, \apj, 596, 768

\bibitem[Spergel et al.\ (2003)]{spergel03} Spergel, D. N., Verde, L., Peiris, H. V.,
     Komatsu, E., Nolta, M. R., Bennett, C. L., Halpern, M., Hinshaw, G., Jarosik,
     N., Kogut, A., Limon, M., Meyer, S. S., Page, L., Tucker, G. S., Weiland, J.
     L., Wollack, E., Wright, E. L. 2003, \apjs, 148, 175

\bibitem[Steidel and Sargent\ (1992)]{SS92} Steidel, C. C., \& Sargent, W. L. W. 
     1992, \apjs, 80, 1

\bibitem[Steidel\  (1993)]{steidel93}
Steidel, C.C. 1993, {\it The Environment and Evolution of
Galaxies}, ed. J.M. Shull \& H.A. Thronson, Jr., 
(Boston: Kluwer Academic Publishers), p. 263

\bibitem[Steidel et al.\ (1999)]{steidel99} Steidel, C. C., Adelberger, K. L., 
     Giavalisco, M., Dickinson, M., \& Pettini, M. 1999, \apj, 519, 1 

\bibitem[Steidel et al.\ (2002)]{steidel02} Steidel, C. C., Kollmeier, J. A., 
     Shapley, A. E., Churchill, C. W., Dickinson, M., \& Pettini, M. 2002, 
     \apj, 570, 526

\bibitem[Strickland \& Stevens\ (2000)]{SS00} Strickland, D. K. \&
Stevens, I. R. 2000, \mnras, 314, 511 

\bibitem[Tytler et al.\ (1987)]{tytler87}		
Tytler, D., Boksenberg, A., Sargent, W.L.W.,
Young, P., \& Kunth, D. 1987, \apjs, 64, 667


\bibitem[Vogt et al.\ (1994)]{vogt} Vogt, S. S., Allen, S. L., Bigelow, B. C., Bresee, L.,
    Brown, B., Cantrall, T., Conrad, A., Coutoure, M., Delaney, C., Epps, H. W., et al. 1994, 
    \procspie, 2198, 362

\end{thebibliography}
\end{document}